\documentclass[aps,twocolumn,nofootinbib,floatfix]{revtex4}
\usepackage{amsfonts}
\usepackage{amssymb}
\usepackage{amsmath}
\usepackage{graphics}
\usepackage{graphicx}
\usepackage{soul}
\usepackage{natbib}
\usepackage{bm}
\usepackage{mathtools}
\usepackage{epstopdf}
\usepackage{color}
\usepackage{lipsum}
\usepackage[colorlinks=true,citecolor=black,linkcolor=black,urlcolor=black,filecolor=black]{hyperref}

\usepackage{acronym}
\newacro{PDF}{probability distribution function}
\newacroplural{PDF}[PDFs]{probability distribution functions}
\newcommand{\PDF}{\ac{PDF}}
\newcommand{\PDFs}{\acp{PDF}}
\newacro{DF}{distribution function}
\newacroplural{DF}[DFs]{distribution functions}
\newcommand{\DF}{\ac{DF}}
\newcommand{\DFs}{\acp{DF}}
\newacro{HMF}{Hamiltonian Mean Field}
\newcommand{\HMF}{\ac{HMF}}

\newcommand{\re}{\mathrm{e}}
\newcommand{\rd}{\mathrm{d}}
\newcommand{\ri}{\mathrm{i}}
\newcommand{\Mtot}{M_{\mathrm{tot}}}
\newcommand{\bw}{\mathbf{w}}
\newcommand{\mP}{\mathcal{P}}
\newcommand{\mPp}{\mathcal{P}^{\prime}}
\newcommand{\mPpp}{\mathcal{P}^{\prime\prime}}
\newcommand{\bvel}{\mathbf{v}}
\newcommand{\bk}{\mathbf{k}}
\newcommand{\deltaD}{\delta_{\mathrm{D}}}
\newcommand{\deltaDp}{\delta_{\mathrm{D}}^{\prime}}
\newcommand{\oF}{\overline{F}}
\newcommand{\ot}{\overline{t}}
\newcommand{\omF}{\overline{\mathcal{F}}}
\newcommand{\bu}{\mathbf{u}}
\newcommand{\mO}{\mathcal{O}}
\newcommand{\dprime}{\prime\prime}
\newcommand{\mF}{\mathcal{F}}
\newcommand{\td}{t_{\mathrm{d}}}
\newcommand{\trelax}{t_{\mathrm{r}}}
\newcommand{\rt}{\mathrm{t}}
\newcommand{\hbM}{\widehat{\mathbf{M}}}
\newcommand{\veps}{\varepsilon}
\newcommand{\beps}{\bm{\varepsilon}}
\newcommand{\bI}{\mathbf{I}}
\newcommand{\oomega}{\overline{\omega}}
\newcommand{\omE}{\overline{\mathcal{E}}}
\newcommand{\Nreal}{N_{\mathrm{real}}}
\newcommand{\Nboot}{N_{\mathrm{boot}}}
\newcommand{\FB}{F_{\mathrm{B}}}
\newcommand{\Qc}{Q_{\mathrm{c}}}
\newcommand{\KI}{K_{\mathrm{I}}}
\newcommand{\KII}{K_{\mathrm{II}}}
\newcommand{\MI}{M_{\mathrm{I}}}
\newcommand{\MII}{M_{\mathrm{II}}}
\newcommand{\vp}{v^{\prime}}

\begin{document}

\title{Kinetic theory of ${1D}$ homogeneous long-range interacting systems
\\
sourced by ${1/N^{2}}$ effects}

\author{Jean-Baptiste Fouvry\footnote{Hubble Fellow}}
\affiliation{Institute for Advanced Study, Princeton, NJ, 08540, USA}
\author{Ben Bar-Or}
\affiliation{Institute for Advanced Study, Princeton, NJ, 08540, USA}
\author{Pierre-Henri Chavanis}
\affiliation{Laboratoire de Physique Th\'eorique, Universit\'e de Toulouse, CNRS, UPS, France}

\begin{abstract}
The long-term dynamics of long-range interacting $N$-body systems
can generically be described by the Balescu-Lenard kinetic equation.
However, for ${1D}$ homogeneous systems, this collision operator
exactly vanishes by symmetry.
These systems undergo a kinetic blocking,
and cannot relax as a whole under ${1/N}$ resonant effects.
As a result, these systems can only relax under ${1/N^{2}}$ effects,
and their relaxation is drastically slowed down.
In the context of the homogeneous Hamiltonian Mean Field model,
we present a new, closed and explicit kinetic equation describing
self-consistently the very long-term evolution of such systems,
in the limit where collective effects can be neglected,
i.e.\ for dynamically hot initial conditions.
We show in particular how that kinetic equation satisfies an $H$--Theorem
that guarantees the unavoidable relaxation to the Boltzmann equilibrium distribution.
Finally, we illustrate how that kinetic equation
quantitatively matches with the measurements from direct $N$-body simulations.
\end{abstract}
\maketitle

\section{Introduction}
\label{sec:Introduction}

The evolution of long-range interacting systems~\citep{CampaDauxois2014}
is generically composed of two stages.
First, the system undergoes a rapid (collisionless) violent relaxation~\citep{LyndenBell1967},
which, owing to strong potential fluctuations, allows for the system to reach
a quasistationary state (i.e.\ a steady state of the Vlasov equation).
Following these drastic orbital rearrangements, the system
has become dynamically frozen for the mean-field dynamics.
It is then only through finite-$N$ effects,
i.e.\ Poisson shot noise due to the finite number of particles,
that the system can keep evolving.
The system undergoes then a slow (collisional)
relaxation that drives it to thermodynamical equilibrium.

The efficiency of the long-term dynamical relaxation of a system therefore depends
on $N$, the system's number of particles.
This dependence is generically accounted for by adding collisional operators
to the collisionless Vlasov equation, i.e.\ by computing the appropriate kinetic equation.
As an example, in the limit where ${1/N}$ effects dominate the dynamics,
such an evolution can generically be described by the Balescu-Lenard equation~\cite{Balescu1960,Lenard1960},
whose generalisation to inhomogeneous systems is only recent~\citep{Heyvaerts2010,Chavanis2012}.
In that context, two body (resonant) encounters lead to a collision operator of order ${ 1/N }$,
so that the relaxation occurs on a timescale scaling like ${ N \td }$,
with $\td$ the system's dynamical time\footnote{One important exception is ${3D}$ self-gravitating systems, where the relaxation time is of order ${ N \td / \log (N) }$, owing to the Coulomb logarithm~\cite{BinneyTremaine2008}.}.

Yet, for ${1D}$ homogeneous systems, such kinetic equations exactly vanish,
i.e.\ two-body encounters are unable to lead to an overall relaxation of the system,
as highlighted in the context of ${1D}$ plasmas~\citep{EldridgeFeix1963,Dawson1964,RouetFeix1991},
the ${1D}$ \HMF\ model~\citep{Zanette2003,Yamaguchi2004,BouchetDauxois2005,Campa2007,Campa2008,RochaFilho2014},
the dynamics of long-range coupled particles on the unit sphere~\cite{Gupta2011,BarreGupta2014,RochaFilho2015,FouvryBarOrChavanis2019},
or even the dynamics of point vortices~\citep{ChavanisLemou2007}.
All these systems are said to being undergoing a kinetic blocking.
For such systems, it is only three-body effects and higher-order correlations that can drive the system's
relaxation to thermodynamical equilibrium~\citep{Bouchet2010,Chavanis2010,Chavanis2011,Sano2012},
making the relaxation time much longer than ${ N \td }$.

As it must originate from perturbations to the system's dynamics of increasing order in ${1/N}$,
it is natural to expect that the timescale for the collisional relaxation of a ${1D}$ homogeneous
system would scale like ${ N^{2} \td }$.
Indeed, this ${ N^{2} \td }$ scaling of the dynamics was observed for ${1D}$ plasmas~\citep{EldridgeFeix1963,Dawson1964,RouetFeix1991},
or for long-range coupled particles on the sphere~\citep{RochaFilho2015,FouvryBarOrChavanis2019}.
In the case of the \HMF\ model, differents scalings proportional to ${ N^{1.7} \td }$~\citep{Zanette2003,Yamaguchi2004},
or even ${ \re^{N} \td }$~\citep{Campa2007} were reported.
But, in~\cite{RochaFilho2014}, these results were convincingly interpreted as being
side effects associated with a too small value of $N$,
and a scaling in $N^{2}$ was recovered through a careful analysis
of simulations with larger values of $N$.

In the present paper, building upon~\cite{RochaFilho2014},
we set out to study such a very long-term dynamics
of the \HMF\ model in the homogeneous limit.
In the limit where collective effects are neglected
(i.e.\ the neglect of the ability of the mean system to amplify perturbations),
we present a new, closed kinetic equation describing
the collisional relaxation of that system on ${ N^{2} \td }$ timescales,
as driven by three-body correlations.
We explore the generic properties of this new collision operator,
and quantitatively compare its predictions to direct $N$-body simulations
of that system.

The paper is organised as follows.
In Section~\ref{sec:KineticEquation}, we briefly present the considered \HMF\ model,
and the kinetic equation describing its relaxation at order ${1/N^{2}}$,
as given by Eq.~\eqref{final_kinetic}.
The detailed procedure followed to obtain that equation is presented
in Appendices~\ref{sec:BBGKY}--\ref{sec:Derivation},
while the effective analytical calculations were performed using a computer algebra system~\citep{MMA}.
In Section~\ref{sec:Properties}, we explore some of the fundamental properties of that kinetic equation,
in particular its well-posedness, its conservation properties,
and its $H$--Theorem that guarantees the relaxation to the homogeneous Boltzmann equilibrium,
provided that it is linearly stable.
Finally, in Section~\ref{sec:Applications}, we quantitatively illustrate how this kinetic equation
matches with direct measurements from numerical simulations,
for hot enough initial distributions.
Finally, we conclude in Section~\ref{sec:Conclusion}.

\section{The kinetic equation}
\label{sec:KineticEquation}

We are interested in the long-term dynamics of the \HMF\ model~\citep{AntoniRuffo1995}.
It is composed of $N$ particles of individual mass ${ \mu = \Mtot / N }$,
with $\Mtot$ the system's total mass.
The canonical phase space coordinates are ${ (\theta , v) }$, and the total Hamiltonian reads
\begin{equation}
H = \frac{1}{2} \sum_{i = 1}^{N} v_{i}^{2} + \mu \sum_{i < j}^{N} U (\theta_{i} - \theta_{j}) ,
\label{total_H_HMF}
\end{equation}
where the pairwise interaction potential is given by
\begin{equation}
U (\theta_{i} - \theta_{j}) = - U_{0} \cos (\theta_{i} - \theta_{j}) ,
\label{def_U}
\end{equation}
with ${ U_{0} > 0 }$ the amplitude of the pairwise coupling.

In the homogeneous limit, the instantaneous statistical state of the system
can be described by the velocity \DF\@, ${ F = F (v , t) }$,
which, following Eq.~\eqref{def_Fn}, is taken to be normalised as
${ \! \int \! \rd \theta \rd v F (v , t) = \Mtot }$.
Describing the long-term relaxation of such a system,
amounts then to describing the long-term evolution of that \DF\@,
as driven by a closed kinetic equation.

When limiting oneself only to ${1/N}$ effects,
the dynamics of that \DF\ is generically given by the Balescu-Lenard equation,
which in the present context reads~\citep{BenettiMarcos2017}
\begin{align}
\frac{\partial F (v_{1})}{\partial t} = & \, \frac{\pi^{2}}{2} U_{0}^{2} \mu \frac{\partial }{\partial v_{1}} \bigg[ \!\! \int \!\! \rd v_{2} \, \!\!\sum_{k = \pm 1} \frac{1}{|\veps_{k} (k v_{1})|^{2}}
\nonumber
\\
\times & \, \deltaD (v_{1} - v_{2}) \, \bigg( \frac{\partial }{\partial v_{1}} - \frac{\partial }{\partial v_{2}} \bigg) \, F (v_{1}) F (v_{2}) \bigg] ,
\label{BL_equation}
\end{align}
where, to shorten the notations,
we do not write explicitly the time-dependence of the \DFs\@.
We also introduced the dielectric coefficient, ${ \veps_{k} (\omega) }$,
explicitly spelled out in Eq.~\eqref{diag_eps}.
As already mentioned in Introduction,
owing to the Dirac delta, ${ \deltaD (v_{1} - v_{2}) }$,
such a collision operator exactly vanishes by symmetry.
The homogeneous \HMF\ system undergoes
a kinetic blocking, and its overall relaxation
is immune to ${1/N}$ two-body effects.

As a result, it is only by being driven by weaker three-body correlations,
associated with ${ 1/N^{2} }$ effects,
that the present system can relax to its thermodynamical equilibrium.
This is the dynamics of interest in this paper.
Following an approach similar to~\cite{RochaFilho2014},
we present in Appendices~\ref{sec:BBGKY}--\ref{sec:Derivation}
our approach to derive such a closed kinetic equation
accounting for ${1/N^{2}}$ effects.

The main steps of this derivation are as follows.
(i) As detailed in Appendix~\ref{sec:BBGKY}, we first derive the usual BBGKY equations,
by obtaining the coupled evolution equations
for the system's $1$-, $2$-, and $3$-body distribution functions.
(ii) As shown in Appendix~\ref{sec:Cluster},
using the cluster expansion,
these coupled evolution equations are written as coupled evolution equations
for the system's $1$-body \DF\@, ${ F (v)}$,
and the $2$-, and $3$-body correlation functions.
The main gain of this rewriting is that these equations
are now sorted by increasing order in ${1/N}$ corrections. 
(iii) As presented in Appendix~\ref{sec:Truncation},
these equations are then truncated at order ${1/N^{2}}$.
At this stage, a key simplification comes from our
neglect of the contributions from collective effects,
i.e.\ the system's ability to amplify perturbations,
as is justified for dynamically hot initial distributions.
(iv) Finally, in Appendix~\ref{sec:Derivation},
we show how this set of (well-posed) coupled
partial differential equations can be solved,
allowing for an explicit and closed expression for the collision operator.
While not intrinsically challenging, such calculations are made cumbersome
because of the large number of terms involved.
These calculations were therefore carried out using \texttt{Mathematica}
and are spelled out in detail in~\cite{MMA}.

All in all, the final kinetic equation derived in that fashion reads
\begin{align}
\frac{\partial F (v_{1})}{\partial t} & \, = \frac{\pi^{3}}{2} \, U_{0}^{4} \, \mu^{2} \frac{\partial }{\partial v_{1}} \bigg[ \mP \!\! \int \!\! \frac{\rd v_{2}}{(v_{1} - v_{2})^{4}}
\nonumber
\\
\times \!\! \int \!\! \rd v_{3} \bigg\{ & \, \deltaD (\bk_{1} \!\cdot\! \bvel) \, \bigg( \bk_{1} \!\cdot\! \frac{\partial }{\partial \bvel} \bigg) F (v_{1}) F(v_{2}) F(v_{3}) 
\nonumber
\\
+ & \, \deltaD (\bk_{2} \!\cdot\! \bvel) \, \bigg( \bk_{2} \!\cdot\! \frac{\partial }{\partial \bvel} \bigg) F (v_{1}) F (v_{2}) F(v_{3}) \bigg\}  \bigg] ,
\label{final_kinetic}
\end{align}
where, to shorten the notations, we introduced
the velocity vector ${ \bvel = (v_{1} , v_{2} , v_{3}) }$,
as well as the two resonance vectors
\begin{equation}
\bk_{1} = (2 , -1 , -1) \;\;\; ; \;\;\; \bk_{2} = (1 , -2 , 1) .
\label{def_bk}
\end{equation}
In that equation, we also introduced $\mP$ as the Cauchy principal value,
which acts on the integral ${ \!\int\! \rd v_{2} }$.
We postpone to Section~\ref{sec:Wellposedness} the justification
of the well-posedness of such a principal value.
Finally, we note that Eq.~\eqref{final_kinetic} is tightly related
to the ${ 1/N^{2} }$ kinetic equation
already put forward in Eq.~{(23)} of~\cite{RochaFilho2014}.
The differences are some corrections in the overall prefactor,
and the sign of the second resonant term.
Compared to~\cite{RochaFilho2014},
in Eq.~\eqref{final_kinetic}, we also performed additional rewritings and manipulations,
that offer a simpler collision operator, involving only one principal value,
and only up to first-order gradients in the system's \DF\@,
as detailed at the end of Appendix~\ref{sec:Derivation}.

As usual, it is possible to rewrite Eq.~\eqref{final_kinetic}
under the form of a continuity equation, reading
\begin{equation}
\frac{\partial F (v_{1})}{\partial t} = \frac{\partial }{\partial v_{1}} \bigg[ \mF (v_{1}) \bigg] ,
\label{def_Flux}
\end{equation}
where the instantaneous flux in velocity space\footnote{With such a convention, the flux is opposite to the direction effectively followed by individual particles during their diffusion.}, ${ \mF (v_{1}) }$,
follows directly from Eq.\eqref{final_kinetic}.

As expected, Eq.~\eqref{final_kinetic} is proportional to ${\mu^{2} \simeq 1/N^{2} }$,
i.e.\ it describes a collisional relaxation on ${ N^{2} \td }$ timescales.
We also note that the collision operator in the r.h.s.\ involves
the system's \DF\ three times, highlighting the fact that this kinetic
equation describes a dynamics sourced by three-body correlations.
These correlations are matched through two different resonance conditions on the velocities,
namely ${ \deltaD (\bk_{1/2} \!\cdot\! \bvel) }$.
We also note that the two resonance terms are opposite one to another,
provided one makes the change ${ (v_{1} \leftrightarrow v_{2}) }$ in the last integrand.
This will prove important to ensure some of the equation's conservation properties,
as detailed in Section~\ref{sec:Conservation}.
Equation~\eqref{final_kinetic} is the main result of this section,
as this closed kinetic equation is the appropriate kinetic equation to describe
the long-term evolution of a dynamically hot ${ 1D }$ homogeneous systems,
sourced by ${ 1/N^{2} }$ effects,
and tailored here to the particular case of the \HMF\ model.
We finally note that Eq.~\eqref{final_kinetic} only holds
as long as the system's mean \DF\ remains linearly stable,
see the end of Section~\ref{sec:Applications} for a more detailed discussion.

\section{Properties}
\label{sec:Properties}

In this section, we now explore some of the key properties
of the kinetic Eq.~\eqref{final_kinetic}.

\subsection{Well-posedness}
\label{sec:Wellposedness}

Given the presence of a high-order resonance denominator in Eq.~\eqref{final_kinetic},
it is not obvious that the kinetic equation is well-defined,
i.e.\ that there are no divergences when ${ v_{2} \to v_{1} }$.
As a result, let us study the behaviour of the integrand in the limit ${ v_{2} \to v_{1} }$.
In order to shorten the notations, we temporarily rewrite Eq.~\eqref{final_kinetic} as
\begin{equation}
\frac{\partial F (v_{1})}{\partial t} = \frac{\pi^{3}}{2} U_{0}^{4} \mu^{2} \frac{\partial }{\partial v_{1}} \bigg[ \mP \!\! \int \!\! \frac{\rd v_{2}}{(v_{1} - v_{2})^{4}} \, K (v_{1} , v_{2}) \bigg] ,
\label{short_rewrite_K}
\end{equation}
where the function ${ K (v_{1} , v_{2}) }$ immediately reads from Eq.~\eqref{final_kinetic}.
Assuming that ${ F (v) }$ is a smooth function, it is straightforward to perform
a limited development of ${ K (v_{1} , v_{1} + \delta v) }$ for ${ \delta v \to 0 }$.
One gets
\begin{equation}
K (v_{1} , v_{1} + \delta v) = K_{3} (v_{1}) \, ( \delta v )^{3} + \mO \big( (\delta v)^{4} \big) ,
\label{DL_K}
\end{equation}
where the first non-zero coefficient, ${ K_{3} (v) }$, reads
\begin{equation}
K_{3} \!=\!  F^{(4)} F F - F^{(3)} F^{\prime} F + 3 F^{\dprime} F^{\prime} F^{\prime} - 3 F F^{\dprime} F^{\dprime} .
\label{def_K3}
\end{equation}
As a consequence, in the vicinity of ${ v_{2} \to v_{1} }$,
the integral from Eq.~\eqref{short_rewrite_K} takes the form
\begin{align}
\mP \!\! \int \!\! \frac{\rd v_{2}}{(v_{1} - v_{2})^{4}} K (v_{1} , v_{2}) & \, \sim \mP \!\! \int \!\! \rd \delta v \,  \frac{K_{3} \, (\delta v)^{3} + \mO \big( (\delta v)^{4}\big)}{(\delta v)^{4}} 
\nonumber
\\
& \, \sim \mP \!\! \int \!\! \rd \delta v \, \bigg\{ \frac{K_{3}}{\delta v} + \mO (1) \bigg\} ,
\label{shape_integral}
\end{align}
which is a meaningful and well-posed integral in terms of a principal value.

\subsection{Boltzmann distribution}
\label{sec:Boltzmann}

The thermodynamical equilibrium states resulting from the collisional relaxation
of a homogeneous $N$-body system are expected to be (shifted) homogeneous
Boltzmann distributions of the form
\begin{equation}
\FB (v) = A \, \re^{- \beta (v - v_{0})^{2}} ,
\label{Boltzmann_DF}
\end{equation}
where $\beta$ is the inverse temperature,
and $A$ a normalisation constant.

Owing to the explicit form of the collision operator from Eq.~\eqref{final_kinetic},
it is straightforward to check that such \DFs\ are indeed equilibrium solutions
of the kinetic equation.
Indeed, noting that the resonance vectors, $\bk_{1}$ and $\bk_{2}$,
from Eq.~\eqref{def_bk} are of zero sum, one has
\begin{align}
\frac{\partial \FB (v_{1})}{\partial t} & \, \propto \bigg\{ \deltaD (\bk_{1} \cdot \bvel) \, (\bk_{1} \cdot \bvel) + \deltaD (\bk_{2} \cdot \bvel) \, (\bk_{2} \cdot \bvel) \bigg\}
\nonumber
\\
& \, = 0 .
\label{collision_FB}
\end{align}
This highlights that the diffusion flux for homogeneous Boltzmann distributions exactly vanishes,
i.e.\ these \DFs\ are equilibrium solutions of the ${1/N^{2}}$ kinetic Eq.~\eqref{final_kinetic}.
In Section~\ref{sec:HTheorem}, owing to an $H$--Theorem,
we will strengthen this result by proving that the homogeneous Boltzmann \DFs\ from Eq.~\eqref{Boltzmann_DF}
are the only equilibrium solutions of the present kinetic equation.

\subsection{Conservation laws}
\label{sec:Conservation}

The kinetic equation~\eqref{final_kinetic} satisfies various conservation properties,
in particular the conservation of
the total mass ${ M(t) }$,
the total momentum, ${ P (t) }$,
and the total energy, ${ E (t) }$,
as we will now briefly justify.
Ignoring numerical prefactors, they are respectively defined as
\begin{align}
M (t) & \, = \!\! \int \!\! \rd v_{1} \, F (v_{1} , t) ,
\nonumber
\\
P (t) & \, = \!\! \int \!\! \rd v_{1} \, v_{1} \, F (v_{1} , t) ,
\nonumber
\\
E (t) & \, = \!\! \int \!\! \rd v_{1} \, \tfrac{1}{2} v_{1}^{2} \, F (v_{1} , t) ,
\label{def_M_P_E}
\end{align}
where the total energy only contains the kinetic energy,
because we assumed that the system remains homogeneous on average.

Following the rewriting from Eq.~\eqref{def_Flux},
the conservation of the total mass follows from the absence
of any boundary contributions, so that
\begin{equation}
\frac{\rd M}{\rd t} = \!\! \int \!\! \rd v_{1} \, \frac{\partial }{\partial v_{1}} \bigg[ \mF (v_{1}) \bigg] = 0.
\label{calc_dMdt}
\end{equation}
A similar calculation can be pursued for the total momentum, and one gets
\begin{align}
\frac{\rd P}{\rd t} & \, = - \!\! \int \!\! \rd v_{1} \, \mF (v_{1}) = 0 ,
\label{calc_dPdt}
\end{align}
using the symmetrisation ${ (v_{1} \leftrightarrow v_{2}) }$ in Eq.~\eqref{final_kinetic}.

Finally, regarding the conservation of energy,
following an integration by parts of Eq.~\eqref{def_M_P_E}, one writes
\begin{equation}
\frac{\rd E}{\rd t} = - \!\! \int \!\! \rd v_{1} \, v_{1} \, \mF (v_{1}) .
\label{calc_dEdt}
\end{equation}
Using the definition of the flux from Eq.~\eqref{def_Flux}, this expression will then involve
an integral of the form ${ \!\int\! \rd v_{1} \rd v_{2} \rd v_{3} }$, which allows us to use
symmetrisations w.r.t.\ the integration variables.
First, we symmetrise all the terms w.r.t.\ the permutation ${ (v_{1} \leftrightarrow v_{2}) }$.
Then, for the subsequent expression, we perform two additional symmetrisations, namely (i)
${ (v_{2} \leftrightarrow v_{3}) }$ for the terms involving the resonance condition ${ \deltaD (\bk_{1} \cdot \bvel) }$, and (ii) ${ (v_{1} \leftrightarrow v_{3}) }$ for the terms involving the resonance condition ${ \deltaD (\bk_{2} \cdot \bvel) }$. By doing so, the resonant denominator from Eq.~\eqref{final_kinetic} remains a sole function ${ (v_{1} \!-\! v_{2}) }$, and this avoids the creation of any other type of resonance conditions.
All these calculations are straightforward and carried out in detail in~\cite{MMA}.
Forgetting prefactors, one gets
\begin{align}
\frac{\rd E}{\rd t} & \, \propto \!\! \int \!\! \rd v_{1} \, \mP \!\! \int \!\! \frac{\rd v_{2}}{(v_{1} - v_{2})^{4}} \!\! \int \!\! \rd v_{3} \, 
\nonumber
\\
\times \bigg\{ & \,  \deltaD (\bk_{1} \cdot \bvel) \, ( \bk_{1} \cdot \bvel ) \bigg( \bk_{1} \cdot \frac{\partial }{\partial \bvel} \bigg) F (v_{1}) F (v_{2}) F (v_{3}) 
\nonumber
\\
+ & \, \deltaD (\bk_{2} \cdot \bvel) \, ( \bk_{2} \cdot \bvel ) \bigg( \bk_{2} \cdot \frac{\partial }{\partial \bvel} \bigg) F (v_{1}) F (v_{2}) F (v_{3})\bigg\}
\nonumber
\\
& \, = 0 ,
\label{calc_dEdt_II}
\end{align}
which exactly vanishes owing to the resonance conditions.

\subsection{$H$--Theorem}
\label{sec:HTheorem}

We define the system's instantaneous entropy as
\begin{equation}
S (t) = - \!\! \int \!\! \rd v_{1} \, s (F (v_{1} , t)) ,
\label{def_S}
\end{equation}
with ${ s (F) = F \ln (F) }$ the entropy functional.
Starting from the rewriting of Eq.~\eqref{def_Flux},
it is straightforward to show that the system's entropy evolves according to
\begin{equation}
\frac{\rd S}{\rd t} = \!\! \int \!\! \rd v_{1} \, \frac{1}{F (v_{1})} \frac{\partial F (v_{1})}{\partial v_{1}} \mF (v_{1}) .
\label{rate_S_init}
\end{equation}
Following Eq.~\eqref{def_Flux}, this expression involves an integral of the form
${ \! \int \! \rd v_{1} \rd v_{2} \rd v_{3} }$,
allowing for symmetrisations w.r.t.\ the integration variables.
We perform the exact same symmetrisations as the one performed in Eq.~\eqref{calc_dEdt_II}
to check for energy conservation.
All these calculations are straightforward, and carried out in detail in~\cite{MMA}.
One gets
\begin{align}
\frac{\rd S}{\rd t} =  \frac{\pi^{3}}{8} U_{0}^{4} \mu^{2} & \, \!\! \int \!\! \rd v_{1} \, \mP \!\! \int \!\! \frac{\rd v_{2}}{(v_{1} - v_{2})^{4}} \!\! \int \!\! \rd v_{3} \, F (v_{1}) F(v_{2}) F(v_{3})
\nonumber
\\
\times \bigg\{ & \, \deltaD (\bk_{1} \cdot \bvel) \, \bigg[ \bk_{1} \cdot \bigg( \frac{F_{1}^{\prime}}{F_{1}} , \frac{F_{2}^{\prime}}{F_{2}} , \frac{F_{3}^{\prime}}{F_{3}} \bigg) \bigg]^{2}
\nonumber
\\
+ & \, \deltaD (\bk_{2} \cdot \bvel) \, \bigg[ \bk_{2} \cdot \bigg( \frac{F_{1}^{\prime}}{F_{1}} , \frac{F_{2}^{\prime}}{F_{2}} , \frac{F_{3}^{\prime}}{F_{3}} \bigg) \bigg]^{2} \bigg\} ,
\label{rate_S}
\end{align}
where we used the shortening notation ${ F_{1} = F (v_{1}) }$,
and ${ F_{1}^{\prime} = \partial F / \partial v_{1} }$.
Given that all the terms involved in this integral are positive, the kinetic Eq.~\eqref{final_kinetic}
satisfies an $H$--Theorem, i.e.\ one has
\begin{equation}
\frac{\rd S}{\rd t} \geq 0 .
\label{H_Theorem}
\end{equation}

The expression of the entropy increase from Eq.~\eqref{rate_S}
allows us then to tackle the question of determining which \DFs\
are equilibrium states for the diffusion, i.e.\ which \DFs\ satisfy ${ \rd S / \rd t = 0 }$.
Provided that one uses the symmetrisation ${ (v_{1} \leftrightarrow v_{2} ) }$, the constraints
associated with the two resonance conditions from Eq.~\eqref{rate_S} are identical,
so that we only need to consider one.
Recalling the expression of the resonance vector $\bk_{1}$ from Eq.~\eqref{def_bk},
and introducing the function ${ G (v) = F^{\prime} (v) / F (v) }$, we note
that a \DF\ is stationary if it satisfies
\begin{equation}
\forall v , \vp \, : \; G \bigg( \frac{v + \vp}{2} \bigg) = \frac{G (v) + G (\vp)}{2} .
\label{constraint_G}
\end{equation}
Because this constraint has to be satisfied for all $v$ and $\vp$,
we can conclude that ${ v \mapsto G (v) }$ has to be a line,
i.e.\ one has
\begin{equation}
G (v) = - 2 \beta (v - v_{0}) \;\;\; \text{ with } \;\;\; \beta > 0 ,
\label{line_G}
\end{equation}
where the constraint ${ \beta > 0 }$
stems from the fact that ${ \!\int\! \rd v F (v) = \Mtot < + \infty }$,
i.e.\ the \DF\ has to be normalised, and cannot get infinitely large for ${ v \to + \infty }$.
Equation~\eqref{line_G} immediately translates to the differential equation
${ F^{\prime} (v) / F (v) = - 2 \beta (v - v_{0}) }$,
which naturally integrates to the (shifted) homogeneous Boltzmann \DF\
introduced in Eq.~\eqref{Boltzmann_DF}.

As a conclusion, the only equilibrium \DFs\ of the kinetic Eq.~\eqref{final_kinetic}
are the (shifted) homogeneous Boltzmann distributions.
This is an important result of this section.
Indeed, while any (stable) \DF\ ${ F(v) }$
was an equilibrium distribution for the ${1/N}$-dynamics
of an homogeneous long-range interacting system,
only homogeneous Boltzmann \DFs\ are equilibrium distributions
for these systems' ${1/N^{2}}$-dynamics.

\subsection{Dimensionless rewriting}
\label{sec:Dimensionless}

In order to have a better grasp at the scalings of Eq.~\eqref{final_kinetic},
let us finally rewrite it under a dimensionless form.

Following the conservation of total energy obtained in Eq.~\eqref{calc_dEdt_II},
we introduce the system's (conserved) velocity dispersion as
\begin{equation}
\sigma^{2} = \frac{1}{\Mtot} \!\! \int \!\! \rd \theta \rd v \, v^{2} \, F (v) .
\label{def_sigma}
\end{equation}
This typical velocity entices us then to define a dimensionless velocity as
${ u = v / \sigma }$,
and a dimensionless time as ${ \ot = t / \td }$ (time) with ${ \td = 1/\sigma }$ the dynamical time.
Similarly, we define the system's dimensionless \PDF\ as
\begin{equation}
\oF (u) = \frac{2 \pi \sigma}{\Mtot} \, F (u \sigma) ,
\label{def_oF}
\end{equation}
that satisfies the normalisation condition ${ \! \int \! \rd u \, \oF (u) = 1 }$.
Finally, in order to assess the ``dynamical temperature'' of the system
and the strength of the associated collective effects,
we introduce the dimensionless stability parameter
\begin{equation}
Q = \frac{2 \sigma^{2}}{U_{0} \Mtot} ,
\label{def_Q}
\end{equation}
following a notation similar to~\citep{Toomre1964}.
The larger $Q$, the more stable the system, and the weaker the collective effects.
In Appendix~\ref{sec:LinearTheory}, we motivate the definition of $Q$,
and directly relate it to the system's dielectric function.

Using these conventions, one can rewrite Eq.~\eqref{final_kinetic} as
\begin{align}
\frac{\partial \oF (u_{1})}{\partial \ot} & \, = \frac{2 \pi}{Q^{4} N^{2}} \frac{\partial }{\partial u_{1}} \bigg[ \mP \!\! \int \!\! \frac{\rd u_{2}}{ (u_{1} - u_{2})^{4} }
\nonumber
\\
\times \!\! \int \!\! \rd u_{3} \, \bigg\{ & \, \deltaD (\bk_{1} \!\cdot\! \bu) \bigg( \bk_{1} \!\cdot\! \frac{\partial }{\partial \bu} \bigg) \oF (u_{1}) \oF (u_{2}) \oF (u_{3})
\nonumber
\\
+ & \, \deltaD (\bk_{2} \!\cdot\! \bu) \bigg( \bk_{2} \!\cdot\! \frac{\partial }{\partial \bu} \bigg) \oF (u_{1}) \oF (u_{2}) \oF (u_{3}) \bigg\} \bigg] ,
\label{final_kinetic_ddim}
\end{align}
where, similarly to Eq.~\eqref{final_kinetic}, we introduced the velocity vector
${ \bu = (u_{1} , u_{2} , u_{3}) }$,
and the resonance vectors, ${ (\bk_{1} , \bk_{2})}$, are given by Eq.~\eqref{def_bk}.
Similarly to Eq.~\eqref{def_Flux}, we can rewrite Eq.~\eqref{final_kinetic_ddim}
under the form of a continuity equation, reading
\begin{equation}
\frac{\partial \oF (u_{1})}{\partial \ot} = \frac{2 \pi}{Q^{4} N^{2}} \frac{\partial }{\partial u_{1}} \bigg[ \omF (u_{1}) \bigg] ,
\label{def_Flux_ddim}
\end{equation}
where the dimensionless instantaneous flux, ${ \omF (u_{1}) }$,
follows from Eq.~\eqref{final_kinetic_ddim}.

Equation~\eqref{final_kinetic_ddim} is the appropriate dimensionless writing
to understand the expected relaxation time of a given system.
Indeed, assuming that the collision operator within brackets is of order unity,
we find therefore that the relaxation time scales like
\begin{equation}
\trelax \simeq Q^{4} \, N^{2} \, \td .
\label{def_Tr}
\end{equation}
It is interesting to note that one recovers that dynamically colder systems,
i.e.\ systems with smaller values of $Q$, relax faster than hotter systems.
However, because collective effects were neglected in the derivation
of Eq.~\eqref{final_kinetic_ddim}, one has to place oneself in the regime ${ Q \gg 1 }$
for the present kinetic equation to apply.
In that dynamically hot regime, collective effects are indeed unimportant,
but, because of the factor $Q^{4}$ in Eq.~\eqref{def_Tr},
relaxation will only occur on very long timescales.

\section{Applications}
\label{sec:Applications}

In order to investigate the validity of the kinetic Eq.~\eqref{final_kinetic},
we now set out to explore numerically the long-term relaxation
of such systems, and compare it with the kinetic prediction.

For clarity, all the details of our numerical implementation
are given in Appendix~\ref{sec:NBody}.
The main difficulty with such a numerical exploration
comes from our neglect of collective effects in the derivation
of the kinetic Eq.~\eqref{final_kinetic}.
As defined in Eq.~\eqref{def_Q}, this asks therefore
for the consideration of initial conditions with ${ Q \gg 1 }$,
for which, following Eq.~\eqref{def_Tr}, the relaxation can only occur on very late timescales,
making the $N$-body simulations more challenging.
The larger $Q$, the weaker the collective effects
(e.g.\@, as can be seen in Fig.~\ref{fig:Nyquist}),
and therefore the better should be the match between the kinetic prediction
and the $N$-body measurements.

\begin{figure}
\begin{center}
\includegraphics[width=0.48\textwidth]{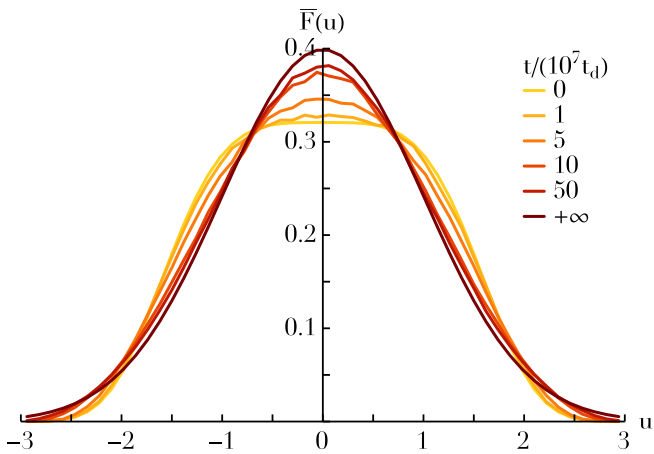}
\caption{Illustration of the overall relaxation of a system's \PDF\@,
${ \oF (u , t) }$, for the non-Gaussian \PDF\ from Eq.~\eqref{def_GaussianAlpha},
for ${ \alpha = 4 }$ and with the dynamical temperature ${ Q = 8.0 }$.
Detailed parameters for these runs are spelled out in Appendix~\ref{sec:NBody}.
Even if such a distribution undergoes a kinetic blocking,
and cannot relax under ${1/N}$ effects,
it is still sensitive to the weaker ${1/N^{2}}$ correlations,
allowing it to slowly relax to the homogeneous Boltzmann thermodynamical equilibrium,
provided that it is linearly stable.
\label{fig:Relaxation}}
\end{center}
\end{figure}

As a first illustration, we present in Fig.~\ref{fig:Relaxation},
an example of a system's relaxation towards equilibrium,
for an initial condition following the non-Gaussian \PDF\
from Eq.~\eqref{def_GaussianAlpha}.
As expected, even if any homogeneous \DF\@, ${ F = F (v) }$,
is submitted to a kinetic blocking, and undergoes no relaxation
through the ${1/N}$ Balescu-Lenard Eq.~\eqref{BL_equation},
it can still relax as a result of higher-order correlation effects,
e.g.\@, as captured by the kinetic Eq.~\eqref{final_kinetic},
whose detailed predictions we may now compute.

\begin{figure}
\begin{center}
\includegraphics[width=0.48\textwidth]{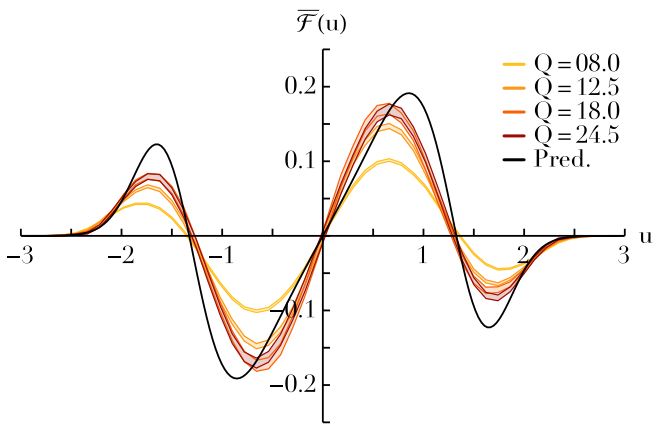}
\caption{Illustration of the dimensionless flux, ${ \omF (u , t \!=\! 0) }$,
as defined in Eq.~\eqref{def_Flux_ddim}
for the non-Gaussian \PDF\ from Eq.~\eqref{def_GaussianAlpha}
with ${ \alpha = 4 }$,
as measured in $N$-body simulations (with the associated errors),
for various initial dynamical temperatures $Q$,
and compared with the prediction from the kinetic Eq.~\eqref{final_kinetic_ddim}.
Detailed parameters for these runs are spelled out in Appendix~\ref{sec:NBody}.
As expected, the larger $Q$, the hotter the system,
and therefore the better the matching with the kinetic prediction
for which collective effects were neglected.
\label{fig:Flux}}
\end{center}
\end{figure}

In Fig.~\ref{fig:Flux}, we illustrate the initial dimensionless flux,
${ \omF (u , t \!=\! 0) }$, as defined in Eq.~\eqref{def_Flux_ddim},
using on the one hand direct measurements from $N$-body simulations
(following the method presented in Appendix~\ref{sec:NBody}),
and on the other hand computing the prediction from the kinetic Eq.~\eqref{final_kinetic_ddim}.
As highlighted, the larger $Q$, the hotter the system,
i.e.\ the weaker the collective effects,
and therefore the better the matching between the $N$-body measurements
and the kinetic prediction.
For systems with smaller velocity dispersions,
Eq.~\eqref{final_kinetic_ddim} does not apply anymore,
and asks to be generalised in order to account for the contribution
of collective effects to hasten or slow down the system's relaxation.
Finally, there are (at least) two possible origins for the slight mismatch
still observed in Fig.~\ref{fig:Flux} between the measured fluxes
and the predicted one:
(i) remaining contributions associated with collective effects,
that are expected to slowly fade out as one increases $Q$;
(ii) some non-vanishing kinetic contributions from the term in ${ G_{2}^{(1)} \!\times\! G_{2}^{(1)} }$ that was neglected in Appendix~\ref{sec:Truncation},
when deriving the system's truncated BBGKY evolution equations.
Such generalisations are beyond the scope of this paper.

\begin{figure}
\begin{center}
\includegraphics[width=0.48\textwidth]{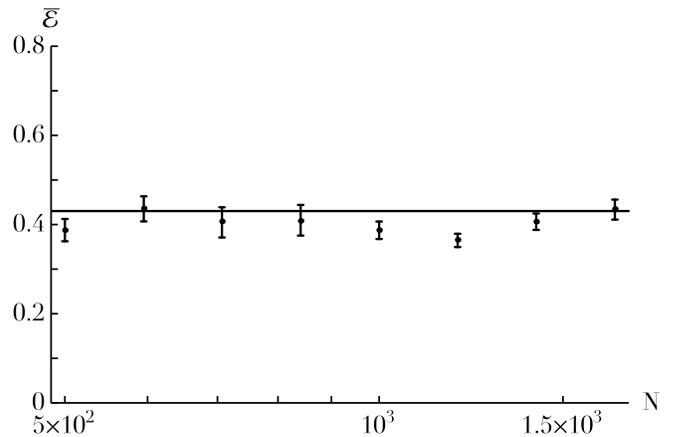}
\caption{Illustration of the dependence of the system's dimensionless relaxation efficiency,
${ \omE = \! \int \! \rd u \, | \omF (u)| }$,
with ${ \omF (u) }$ the dimensionless flux as defined in Eq.~\eqref{def_Flux_ddim},
for the same initial conditions as in Fig.~\ref{fig:Flux},
with ${ Q = 24.5 }$ and various values of $N$.
The black line is the prediction from the kinetic Eq.~\eqref{final_kinetic_ddim},
while the gray dots are the $N$-body measurements,
with the associated errors.
Detailed parameters for these runs are spelled out in Appendix~\ref{sec:NBody}.
As expected from the scaling of Eq.~\eqref{def_Flux_ddim},
the dimensionless relaxation efficiency is independent of $N$.
\label{fig:Nscaling}}
\end{center}
\end{figure}

As can be noted from the overall prefactor in Eq.~\eqref{final_kinetic_ddim},
one expects the timescale for the system's relaxation to scale like ${ N^{2} \td }$,
w.r.t.\ $N$ the number of particles.
This is investigated in Fig.~\ref{fig:Nscaling},
where we illustrate the dependence of the system's relaxation efficiency,
${ \omE = \! \int \! \rd u \, |\omF (u)| }$,
as a function of the number of particles.
In that figure, we observe that the dimensionless relaxation efficiency
is independent of $N$,
so that the scaling of the relaxation time from Eq.~\eqref{def_Tr}
is indeed recovered.

Before concluding, let us finally briefly describe
the system's possible dynamics depending
on the value of the dynamical temperature $Q$.
As derived in Eq.~\eqref{stability_Q},
we have shown that for ${ Q \leq \Qc }$,
the homogeneous system is initially linearly unstable.
This then leads to the following possible behaviours.

(i) For ${ Q \leq \Qc }$, the homogeneous system is initially linearly unstable,
so that it rapidly becomes inhomogeneous.
For such an inhomogeneous configuration,
the kinetic blocking from Eq.~\eqref{BL_equation} does not hold anymore.
Provided that the system remains inhomogeneous,
its dynamics is driven by the inhomogeneous Balescu-Lenard equation,
and drives a relaxation,
whose associated relaxation time scales linearly with $N$~\citep{BenettiMarcos2017}.
As noted in~\cite{Campa2008},
an interesting case is given by an initial configuration
satisfying ${ 1 < Q < \Qc }$.
While such an initial condition is initially unstable,
its final homogeneous Boltzmann equilibrium is stable (since ${ 1 < Q }$).
As a consequence, at some point,
the system has to evolve from a
inhomogeneous distribution to a homogeneous one,
which accordingly delays the relaxation.

(ii) For ${ Q \gtrsim \Qc }$, because it is linearly stable,
the system remains initially homogeneous,
and evolves according to a homogeneous kinetic equation in ${ 1/N^{2} }$.
However, because it is close to the stability threshold,
collective effects, i.e.\ the system's ability to amplify perturbations,
have to be taken into account.
This asks for a kinetic equation more general than Eq.~\eqref{final_kinetic_ddim},
where we neglected collective effects.
As noted in~\cite{Campa2008}, an interesting case
is given by an initial configuration such that ${ \Qc < Q < 1 }$.
Such a system is initially stable so that it will first remain homogeneous
and undergo a slow ${ 1/N^{2} }$ relaxation.
Yet, since ${ Q < 1 }$, the associated homogeneous Boltzmann equilibrium
is unstable.
As a consequence during its (homogeneous) relaxation,
the system will unavoidably become unstable at some point.
This will drive a dynamical phase transition rapidly making it inhomogeneous.
The final stages of the relaxation are then the ones
of an inhomogeneous relaxation, that scales in ${ 1/N }$.
Owing to this dynamical phase transition, one expects therefore
the system's overall relaxation time to have an intermediate
scaling between $N$ and $N^{2}$.

(iii) Finally, for ${ Q \gg \Qc }$, one recovers
the case considered in the present paper.
Because it is dynamically so hot, the system will not undergo any instability.
It will therefore remain homogeneous throughout its relaxation.
On the same grounds, collective effects are unimportant and can be neglected.
The system's dynamics is therefore described
by Eq.~\eqref{final_kinetic_ddim},
and leads to a relaxation time scaling like $N^{2}$,
as illustrated in Fig.~\ref{fig:Nscaling}.

\section{Conclusion}
\label{sec:Conclusion}

In the present paper, we focused our attention on the description of the very long-term
dynamics of the \HMF\ model in the homogeneous limit,
one particular example of a ${1D}$ long-range interacting system.
As highlighted in Eq.~\eqref{BL_equation},
such systems are generically submitted to a kinetic blocking
that prevents their relaxation as a whole under ${1/N}$ resonant effects.
As such, their evolution is drastically slowed down,
and is only made possible by the cumulative contributions
of higher-order ${1/N^{2}}$ effects.

Placing ourselves within the dynamically hot limit,
for which collective effects can be neglected,
and following an approach similar to~\cite{RochaFilho2014},
we showed how one could explicitly solve the BBGKY hierarchy of equations,
truncated at order ${1/N^{2}}$.
This led us to Eq.~\eqref{final_kinetic}, a closed, explicit,
and self-consistent kinetic equation describing
the long-term relaxation of the system's homogeneous \DF\
as driven by ${1/N^{2}}$ effects,
as long as the mean system remains linearly stable.

We put forward the main properties of that new kinetic equation,
in particular the fact that it satisfies an $H$--Theorem,
that guarantees the unavoidable relaxation of the system
towards the homogeneous Boltzmann thermodynamical equilibrium
(provided that it is linearly stable).
This result highlights the fundamental importance
of Boltzmann's $H$--Theorem that keeps being satisfied
in the present ${1/N^2}$ context
where the relaxation is sourced by the product of three \DFs\@.
This therefore further extends the validity of Boltzmann's $H$--Theorem
beyond the traditional kinetic equations such as the collisional Boltzmann,
Landau and Balescu-Lenard equations,
which are sourced only by product of two \DFs\@.

In Eq.~\eqref{def_Tr}, we subsequently detailed how the present formalism
predicts a relaxation time scaling like ${ \trelax \simeq Q^{4} N^{2} \td }$,
with $\td$ the dynamical time,
and $Q$ the system's stability parameter, as defined in Eq.~\eqref{def_Q}.
In particular, this implied a relaxation time scaling like $N^{2}$
w.r.t.\ the number of particles,
a scaling already thoroughly checked in~\cite{RochaFilho2014}.

Finally, in Section~\ref{sec:Applications}, we presented 
explicit comparisons of this new kinetic equation
with numerical measurements from direct $N$-body simulations.
We illustrated in Fig.~\ref{fig:Relaxation} how at this ${ 1/N^{2} }$ order,
the system does not suffer anymore from a kinetic blocking,
and can indeed relax to the homogeneous Boltzmann equilibrium,
provided that it is linearly stable.
We quantitatively showed in Fig.~\ref{fig:Flux}
how the numerically measured diffusion fluxes converge
to the kinetic prediction, as the system is made hotter
so that collective effects become more and more negligible.
We also illustrated in Fig.~\ref{fig:Nscaling} how the $N^{2}$
scaling of the relaxation time is also recovered numerically.

The kinetic equation presented in Eq.~\eqref{final_kinetic}
is only a first step towards the detailed characterisation of the (very)
long-term dynamics of long-range interacting systems.
In the present context, calculations were made more  tractable
through the following assumptions:
(i) the \HMF\ model contains only one harmonic, ${ k = \pm 1 }$,
in its pairwise interaction, reducing drastically the allowed resonances;
(ii) we neglected contributions associated with collective effects,
which prevented us from having to solve and include the linear
response theory of the system;
(iii) finally, in the evolution equation for the three-body correlation function,
${ \partial G_{3} / \partial t }$, we neglected the contributions from
the source term in ${ G_{2}^{(1)} \!\times\! G_{2}^{(1)} }$
in Eq.~\eqref{BBGKY_G3},
that would have led to an additional collision term proportional to $F^{4}$,
instead to $F^{3}$ for the dominant term included in Eq.~\eqref{final_kinetic}.
Further work should try to alleviate these shortcomings,
by allowing for more complex resonances,
accounting for collective effects to describe dynamically colder systems
that are linearly more responsive,
and finally by accounting for possible contributions from higher order terms
in the system's \DF\@.

More generally, one should investigate the structure
of the collision operators for even higher-order kinetic equations,
e.g.\@, at order ${ 1/N^{3} }$. For example, as recovered in the classical
${1/N}$ Landau and Balescu-Lenard equations, and as recovered here
for the ${ 1/N^2 }$ kinetic equation, Boltzmann distributions are always
found, a posteriori, to be equilibrium states of the collision operator.
It would be of interest to investigate whether or not
such a property generically holds for higher order expansions,
and, if so, understand why.

\begin{acknowledgments}
JBF acknowledges support from Program number HST-HF2-51374 which was provided
by NASA through a grant from the Space Telescope Science Institute, which is
operated by the Association of Universities for Research in Astronomy,
Incorporated, under NASA contract NAS5--26555.  BB is supported by membership
from Martin A. and Helen Chooljian at the Institute for Advanced Study.
\end{acknowledgments}

\appendix

\section{The BBGKY hierarchy}
\label{sec:BBGKY}

In this Appendix, we briefly repeat the derivation of the BBGKY hierarchy,
to describe the dynamics of a long-range coupled $N$-body system.
Notations and normalisations are inspired from the ones considered in~\cite{Balescu1997}.

We assume that the system is composed of $N$ identical particles
of individual mass ${ \mu = \Mtot / N }$, with $\Mtot$ the system's total mass.
We introduce the system's $N$-body \PDF\@, ${ P_{N} (\bw_{1} , ... , \bw_{N} , t) }$,
with ${ \bw = (\theta , v) }$ the phase space coordinates, normalised so that ${ \!\int\! \rd \bw_{1} ... \rd \bw_{N} P_{N} = 1 }$.
The dynamics of $P_{N}$ is governed by Liouville's equation
\begin{equation}
\frac{\partial P_{N}}{\partial t} + \bigg[ P_{N} , H_{N} \bigg]_{N} = 0 ,
\label{Liouville_PN}
\end{equation}
where we introduced the full $N$-body Hamiltonian
\begin{equation}
H_{N} (\bw_{1} , ... , \bw_{N}) = \frac{1}{2} \sum_{i = 1}^{N} v_{i}^{2} + \mu \sum_{ i < j }^{N} U (\theta_{i} - \theta_{j}) ,
\label{def_HN}
\end{equation}
with ${ U (\theta_{i} - \theta_{j}) }$ the considered pairwise interaction.
In Eq.~\eqref{Liouville_PN}, we also introduced the Poisson bracket over $N$
particles as
\begin{equation}
\bigg[ P_{N} , H_{N} \bigg]_{N} = \sum_{i = 1}^{N} \bigg\{ \frac{\partial P_{N}}{\partial \theta_{i}} \frac{\partial H_{N}}{\partial v_{i}} - \frac{\partial P_{N}}{\partial v_{i}} \frac{\partial H_{N}}{\partial \theta_{i}} \bigg\} .
\label{def_Poisson_N}
\end{equation}
We can subsequently define the system's reduced \PDFs\ as
\begin{equation}
P_{n} (\bw_{1} , ... , \bw_{n} , t) = \!\! \int \!\! \rd w_{n+1} ... \rd  w_{N} \, P_{N} (\bw_{1} , ... , \bw_{N} , t) .
\label{def_Pn}
\end{equation}
Integrating Eq.~\eqref{Liouville_PN} w.r.t. all particles but the $n$ first, we obtain the BBGKY hierarchy of equations, namely
\begin{equation}
\frac{\partial P_{n}}{\partial t} + \bigg[ P_{n} , H_{n} \bigg]_{n} + (N - n) \!\! \int \!\! \rd \bw_{n+1} \, \bigg[ P_{n+1} , \mu \, \delta H_{n+1} \bigg]_{n} = 0 ,
\label{BBGKY_Pn}
\end{equation}
where we used the symmetry of $P_{N}$ w.r.t.\ exchanges of particles.
Similarly to Eq.~\eqref{def_HN}, the $n$-body Hamiltonian $H_{n}$
(resp. ${ [ \,\cdot\, , \,\cdot\, ]_{n} }$ the Poisson bracket over $n$ particles)
naturally follows from Eq.~\eqref{def_HN} (resp. Eq.~\eqref{def_Poisson_N}),
provided that one replaces $N$ by $n$.
In Eq.~\eqref{BBGKY_Pn}, we also introduced ${ \delta H_{n+1} }$
as the specific interaction energy of the ${ (n+1)^{\mathrm{th}} }$ particle
with the $n$ first particles. It reads
\begin{equation}
\delta H_{n+1} (\bw_{1} , ... , \bw_{n+1}) = \sum_{i = 1}^{n} U (\theta_{i} - \theta_{n+1}) .
\label{def_deltaH}
\end{equation}
As usual, we note that the BBGKY hierarchy from Eq.~\eqref{BBGKY_Pn}
is not closed, as the evolution equation for ${ \partial P_{n} / \partial t }$
involves the higher-order \PDF\@, $P_{n+1}$.

In order to simplify the combinatorial prefactors appearing in Eq.~\eqref{BBGKY_Pn},
we finally introduce the reduced \DFs\@, $F_{n}$, as
\begin{equation}
F_{n} = \mu^{n} \frac{N!}{(N - n)!} \, P_{n} .
\label{def_Fn}
\end{equation}
With such a choice, these \DFs\ scale as ${ F_{n} \sim 1 }$,
w.r.t.\ $N$ the total number of particles.
We can then rewrite the BBGKY hierarchy Eq.~\eqref{BBGKY_Pn} under the simple form
\begin{equation}
\frac{\partial F_{n}}{\partial t} + \bigg[ F_{n} , H_{n} \bigg]_{n} + \!\! \int \!\! \rd \bw_{n+1} \, \bigg[ F_{n + 1} , \delta H_{n + 1} \bigg]_{n} = 0 .
\label{BBGKY_Fn}
\end{equation}
The three first equations of the BBGKY hierarchy,
i.e.\ the evolution equations for $F_{1}$, $F_{2}$, and $F_{3}$,
will be the starting point of the derivation of the kinetic equation
presented in Eq.~\eqref{final_kinetic}.

\section{The cluster expansion}
\label{sec:Cluster}

In Appendix~\ref{sec:BBGKY}, we briefly rederived the BBGKY hierarchy
of evolution equations for the system's reduced \DFs\@.
In order to be able to perform perturbative developments w.r.t.\ $N$
the total number of particles, we now introduce the cluster representation
of the \DFs\@, following an approach similar to the one presented in~\cite{Balescu1997}.

We introduce the system's $2$-body correlation function, ${ G_{2} (\bw_{1} , \bw_{2}) }$,
as
\begin{align}
F_{2} (1,2) = & \, F_{1} (1) \, F_{1} (2) + G_{2} (1,2) ,
\label{def_G2}
\end{align}
where we used the shortened notation ${ F_{1} (1) = F_{1} (\bw_{1}) }$.
This correlation function characterises how much the statistics of the distribution of $2$ particles
differs from being separable.
Similarly, we introduce the system's $3$-body correlation function, ${ G_{3} (\bw_{1} , \bw_{2} , \bw_{3}) }$, as
\begin{align}
F_{3} (1,2,3) & \, = F_{1} (1) F_{1} (2) F_{1} (3)
\nonumber
\\
+ & \, F_{1} (1) G_{2} (2 , 3) \!+\! F_{1} (2) G_{2} (1,3) \!+\! F_{1} (3) G_{2} (1,2)
\nonumber
\\
+ & \, G_{3} (1,2,3) .
\label{def_G3}
\end{align}
Finally, we introduce the $4$-body correlation function, ${ G_{4} (\bw_{1} , \bw_{2} , \bw_{3} , \bw_{4}) }$, as
\begin{align}
& \, F_{4} (1,2,3,4) = F_{1} (1) F_{1} (2) F_{1} (3) F_{1} (4)
\nonumber
\\
+ & \, \bigg\{ F_{1} (1) F_{1} (2) G_{2} (3,4) \!+\! F_{1} (1) F_{1} (3) G_{2} (2 , 4)
\nonumber
\\
& \!\!  + F_{1} (1) F_{1} (4) G_{2} (2 , 3) \!+\! F_{1} (2) F_{1} (3) G_{2} (1,4)
\nonumber
\\
& \!\! + F_{1} (2) F_{1} (4) G_{2} (1,3) \!+\! F_{1} (3) F_{1} (4) G_{2} (1,2) \bigg\}
\nonumber
\\
+ & \, G_{2} (1,2) G_{2} (3,4) \!+\! G_{2} (1,3) G_{2} (2,4) \!+\! G_{2} (1,4) G_{2} (2,3)
\nonumber
\\
+ & \, \bigg\{ F_{1} (1) G_{3} (2,3,4) \!+\! F_{1} (2) G_{3} (1,3,4)
\nonumber
\\
& \!\! + F_{1} (3) G_{3} (1,2,4) \!+\! F_{1} (4) G_{3} (1,2,3) \bigg\}
\nonumber
\\
+ & \, G_{4} (1,2,3,4) . 
\label{def_G4}
\end{align}
The best way to check for the sanity of the previous definitions is to compute the normalisation of the correlation functions, and their scaling w.r.t.\ $N$.
Integrating Eqs.~\eqref{def_G2},~\eqref{def_G3} and~\eqref{def_G4} w.r.t.\ their phase space coordinates, one obtains
\begin{align}
& \, \!\! \int \!\! \rd 1 \, F_{1} (1) = \mu N \sim 1 ,
\nonumber
\\
& \, \!\! \int \!\! \rd 1 \, \rd 2 \, G_{2} (1,2) = - \mu^{2} N \sim \frac{1}{N} ,
\nonumber
\\
& \, \!\! \int \!\! \rd 1 \, \rd 2 \, \rd 3 \, G_{3} (1,2,3) = 2 \mu^{3} N \sim \frac{1}{N^{2}} ,
\nonumber
\\
& \, \!\! \int \!\! \rd 1 \, \rd 2 \, \rd 3 \, \rd 4 \, G_{4} (1,2,3,4) = - 6 \mu^{4} N \sim \frac{1}{N^{3}} ,
\label{normalisations_G}
\end{align}
where we used the shortening notation ${ \rd 1 = \rd \bw_{1} }$.
Owing to these scalings, one can therefore use the correlation functions
to perform perturbative expansions w.r.t. the small parameter ${ 1/N }$.

The next step of the calculation is now to inject the previous decompositions
into the three first equations of the BBGKY hierarchy, as given by Eq.~\eqref{BBGKY_Fn},
in order to obtain the evolution equations for ${ \partial F_{1} / \partial t }$, ${ \partial G_{2} / \partial t}$
and ${ \partial G_{3} / \partial t }$.
Such equations can be cumbersome to derive, and were obtained using computer algebra in~\cite{MMA}.
Writing the system's $1$-body \DF\ as ${ F = F_{1} }$, its time evolution is given by
\begin{align}
& \, \frac{\partial F (1)}{\partial t}
\nonumber
\\
+ & \, v_{1} \, \frac{\partial F (1)}{\partial \theta_{1}}
\nonumber
\\
- & \, \frac{\partial F (1)}{\partial v_{1}} \, \!\! \int \!\! \rd 2 \, F (2) \,U' (\theta_{1} \!-\! \theta_{2})
\nonumber
\\
- & \, \!\! \int \!\! \rd 2 \, \frac{\partial G_{2} (1,2)}{\partial v_{1}} \, U' (\theta_{1} \!-\! \theta_{2})
\nonumber
\\
& \, = 0 .
\label{BBGKY_F}
\end{align}
The second equation of the hierarchy, for ${ \partial G_{2} / \partial t }$, reads
\begin{align}
& \, \frac{\partial G_{2} (1,2)}{\partial t}
\nonumber
\\
+ \, \bigg[ & \, v_{1} \, \frac{\partial G_{2} (1,2)}{\partial \theta_{1}}
\nonumber
\\
& \, - \frac{\partial G_{2} (1,2)}{\partial v_{1}} \!\! \int \!\! \rd 3 \, F (3) \, U' (\theta_{1} \!-\! \theta_{3})
\nonumber
\\
& \, - \frac{\partial F (1)}{\partial v_{1}} \!\! \int \!\! \rd 3 \, G_{2} (2 , 3) \, U' (\theta_{1} \!-\! \theta_{3})
\nonumber
\\
& \, - \mu \, \frac{\partial F (1)}{\partial v_{1}} \, F (2) \, U' (\theta_{1} \!-\! \theta_{2})
\nonumber
\\
& \, - \mu \, \frac{\partial G_{2} (1,2)}{\partial v_{1}} \, U' (\theta_{1} \!-\! \theta_{2})
\nonumber
\\
& \, - \!\! \int \!\! \rd 3 \, \frac{\partial G_{3} (1,2,3)}{\partial v_{1}} \, U' (\theta_{1} \!-\! \theta_{3})
\nonumber
\\
& \, \bigg]_{(1,2)} = 0 ,
\label{BBGKY_G2}
\end{align}
where we introduced the symmetrising notation
\begin{equation}
\bigg[ G (1,2) \bigg]_{(1,2)} = G (1,2) + G (2,1) .
\label{symmetrisation_2}
\end{equation}
Finally, the third equation of the hierarchy, for ${ \partial G_{3} / \partial t }$, reads
\begin{align}
& \, \frac{\partial G_{3} (1,2,3)}{\partial t}
\nonumber
\\
& \, + \, \bigg[  v_{1} \frac{\partial G_{3} (1,2,3)}{\partial \theta_{1}}
\nonumber
\\
& \, - \frac{\partial G_{3} (1,2,3)}{\partial v_{1}} \!\! \int \!\! \rd 4 \, F (4) \, U' (\theta_{1} \!-\! \theta_{4})
\nonumber
\\
& \, - \frac{\partial F(1)}{\partial v_{1}} \!\! \int \!\! \rd 4 \, G_{3} (2,3,4) \, U' (\theta_{1} \!-\! \theta_{4})
\nonumber
\\
& \, - \, \mu \, \frac{\partial F (1)}{\partial v_{1}} \, G_{2} (2 , 3) \, \big\{ U' (\theta_{1} \!-\! \theta_{2}) + U' (\theta_{1} \!-\! \theta_{3}) \big\}
\nonumber
\\
& \, - \mu \, \frac{\partial G_{2} (1 , 2)}{\partial v_{1}} \, F (3) \, U' (\theta_{1} \!-\! \theta_{3}) 
\nonumber
\\
& \, - \mu \, \frac{\partial G_{2} (1,3)}{\partial v_{1}} \, F (2) \, U' (\theta_{1} \!-\! \theta_{2})
\nonumber
\\
& \, - \frac{\partial G_{2} (1,2)}{\partial v_{1}} \!\! \int \!\! \rd 4 \, G_{2} (3 ,4) \, U' (\theta_{1} \!-\! \theta_{4})
\nonumber
\\
& \, - \frac{\partial G_{2} (1,3)}{\partial v_{1}} \!\! \int \!\! \rd 4 \, G_{2} (2,4) \, U' (\theta_{1} \!-\! \theta_{4})
\nonumber
\\
& \, - \mu \, \frac{\partial G_{3} (1,2,3)}{\partial v_{1}} \, \big\{ U' (\theta_{1} \!-\! \theta_{2}) + U' (\theta_{1} \!-\! \theta_{3}) \big\}
\nonumber
\\
& \, - \!\! \int \!\! \rd 4 \, \frac{\partial G_{4} (1,2,3,4)}{\partial v_{1}} \, U' (\theta_{1} \!-\! \theta_{4})
\nonumber
\\
& \, \bigg]_{(1,2,3)} = 0 .
\label{BBGKY_G3}
\end{align}
Here, similarly to Eq.~\eqref{symmetrisation_2}, we introduced the symmetrising notation
\begin{equation}
\bigg[ G (1,2,3) \bigg]_{(1,2,3)} \!\!\!\!\!\!\! = G (1,2,3) \!+\! G (2,3,1) \!+\! G (3,1,2) ,
\label{symmetrisation_3}
\end{equation}
where it is assumed that the function $G$ is symmetric w.r.t.\ its two last indices,
i.e.\ one should have ${ G (1,2,3) = G (1,3,2) }$, as is the case for the term in Eq.~\eqref{BBGKY_G3}.

Equations~\eqref{BBGKY_F},~\eqref{BBGKY_G2} and~\eqref{BBGKY_G3}
are the starting blocks to obtain a self-consistent set of coupled evolution equations
describing the system's entire dynamics up to order ${ 1/N^{2} }$.

\section{Truncating the BBKGY hierarchy}
\label{sec:Truncation}

In this Appendix, we detail how one may truncate
Eqs.~\eqref{BBGKY_F},~\eqref{BBGKY_G2}, and~\eqref{BBGKY_G3}
to lay the groundwork to derive the closed kinetic equation presented in Eq.~\eqref{final_kinetic}.

The first step of these simplifications is to perform a truncation at order ${1/N^{2}}$
of these three evolution equations.
In Eq.~\eqref{BBGKY_F}, we also note that the collision term for ${ \partial F / \partial t }$ only involves ${ G_{2} (1,2) }$, whose norm scales like ${ 1/N }$,
as given by Eq.~\eqref{normalisations_G}.
As a consequence, if one aims at deriving a kinetic equation at order ${ 1N^{2} }$,
it is essential to account for the corrections of order ${ 1/N^{2} }$
that can arise in $G_{2}$.
To perform the truncation at order ${ 1/N^{2} }$,
we therefore introduce explicitly the small parameter ${ \veps = 1/N }$.
Following the definition ${ \mu = \Mtot / N }$, and the scalings from Eq.~\eqref{normalisations_G}, we perform the replacements
\begin{align}
& \, \mu \to \veps \mu \; ; \; G_{2} \to \veps G_{2}^{(1)} + \veps^{2} G_{2}^{(2)}
\nonumber
\\
& \, G_{3} \to \veps^{2} G_{3} \; ; \; G_{4} \to \veps^{3} G_{4} .
\label{replacements_scaling}
\end{align}
Using this rewriting, we then keep in the evolution equations only
terms up to order $\veps^{2}$.
Moreover, owing to the split of $G_{2}$ in two components,
we can split the associated evolution Eq.~\eqref{BBGKY_G2}
in two components, namely at order ${ 1/N }$ (resp.\ ${1/N^{2}}$)
that will govern the dynamics of ${ \partial G_{2}^{(1)} / \partial t }$
(resp.\ ${ \partial G_{2}^{(2)} / \partial t }$).

A subsequent simplification arises from the homogeneous assumption,
i.e.\ the assumption that system's mean \DF\ remains a function of $v$ only.
This allows us to get rid of the phase mixing term, ${ v_{1} \partial F / \partial \theta_{1} }$, in Eq.~\eqref{BBGKY_F},
and also get rid of all the mean-field potential components,
i.e.\ terms involving ${ \! \int \! \rd 2 F(2) \, U' (\theta_{1} \!-\! \theta_{2}) = 0 }$.

In order to ease the derivations of the kinetic equation,
we also assume that the mean system is sufficiently dynamically hot
for collective effects to be negligible\footnote{In the context of ${1/N}$ dynamics,
such an assumption gets the Balescu-Lenard equation to reduce to
the Landau kinetic equation~\citep{Heyvaerts2010,Chavanis2012}.}.
Such an assumption amounts to neglecting the backreaction of a correlation
function on the perturbating potential in which that same correlation function is evolving.
As a result, we perform the following simplifications
\begin{align}
& \, \text{For } \frac{\partial G_{2}^{(1)} (1,2)}{\partial t} \text{: } \!\! \int \!\! \rd 3 \, G_{2}^{(1)} (2,3) \, U' (\theta_{1} \!-\! \theta_{3}) \to 0 ,
\nonumber
\\
& \, \text{For } \frac{\partial G_{2}^{(2)} (1,2)}{\partial t} \text{: } \!\! \int \!\! \rd 3 \, G_{2}^{(2)} (2,3) \, U' (\theta_{1} \!-\! \theta_{3}) \to 0 ,
\nonumber
\\
& \, \text{For } \frac{\partial G_{3} (1,2,3)}{\partial t} \text{: } \!\! \int \!\! \rd 4 \, G_{3} (2,3,4) \, U' (\theta_{1} \!-\! \theta_{4}) \to 0 .
\label{simplification_collective}
\end{align}

Finally, we perform three last approximations:
(i) in the evolution equation for ${ \partial F / \partial t }$,
we may neglect the contribution from ${ G_{2}^{(1)} }$
that is responsible for the usual ${1/N}$ Landau equation,
which identically vanishes for ${1D}$ homogeneous systems,
(as highlighted in Eq.~\eqref{BL_equation});
(ii) in the evolution equation for ${ \partial G_{2}^{(2)} / \partial t }$,
we may safely neglect the contributions from the source term
proportional to ${ \mu \partial G_{2}^{(1)} (1,2) / \partial v_{1} }$,
as one can check that it does not contribute to the kinetic equation
(see~\cite{MMA});
(iii) in the evolution equation for ${ \partial G_{3} / \partial t }$,
we neglect the contributions from the term proportional to
${ G_2^{(1)} \!\times\! G_{2}^{(1)} }$, as it will lead to a collision operator
proportional to $F^{4}$,
while the other source terms will lead to a collision operator
proportional to $F^{3}$,
that dominates for sufficiently dynamically hot systems.

Following these various truncations and simplifications,
we now have at our disposal a set of four coupled evolution equations,
that jointly describe the long-term dynamics of the considered system
at order ${ 1 / N^{2} }$.
The dynamics of ${ \partial F (1) / \partial t }$ reads
\begin{align}
& \, \frac{\partial F(1)}{\partial t}
\nonumber
\\
& \, - \!\! \int \!\! \rd 2 \, \frac{\partial G_{2}^{(2)} (1,2)}{\partial v_{1}} \, U' (\theta_{1} \!-\! \theta_{2})
\nonumber
\\
& \, = 0 .
\label{short_evol_F}
\end{align}
The dynamics of ${ \partial G_{2}^{(1)} (1,2) / \partial t }$ is given by
\begin{align}
& \, \frac{\partial G_{2}^{(1)} (1,2)}{\partial t}
\nonumber
\\
& \, + \bigg[ v_{1} \, \frac{\partial G_{2}^{(1)} (1,2)}{\partial \theta_{1}}
\nonumber
\\
& \, - \mu \frac{\partial F (1)}{\partial v_{1}} \, F (2) \, U' (\theta_{1} \!-\! \theta_{2})
\nonumber
\\
& \,  \bigg]_{(1,2)} = 0 ,
\label{short_evol_G21}
\end{align}
while the dynamics of the second-order correction,
${ \partial G_{2}^{(2)} (1,2) / \partial t }$ takes the form
\begin{align}
& \, \frac{\partial G_{2}^{(2)} (1,2)}{\partial t}
\nonumber
\\
& \, + \bigg[ v_{1} \, \frac{\partial G_{2}^{(2)} (1,2)}{\partial \theta_{1}}
\nonumber
\\
& \, - \!\! \int \!\! \rd 3 \, \frac{\partial G_{3} (1,2,3)}{\partial v_{1}} \, U' (\theta_{1} \!-\! \theta_{3})
\nonumber
\\
& \,  \bigg]_{(1,2)} = 0 .
\label{short_evol_G22}
\end{align}
Finally, the dynamics of the $3$-body correlation function,
${ \partial G_{3} (1,2,3)/\partial t }$, reads
\begin{align}
& \, \frac{\partial G_{3} (1,2,3)}{\partial t}
\nonumber
\\
& \, + \bigg[ v_{1} \, \frac{\partial G_{3} (1,2,3)}{\partial \theta_{1}}
\nonumber
\\
& \, - \mu \frac{\partial F (1)}{\partial v_{1}} \, G_{2}^{(1)} (2,3) \, \big\{ U' (\theta_{1} \!-\! \theta_{2}) + U' (\theta_{1} \!-\! \theta_{3}) \big\}
\nonumber
\\
& \, - \mu \frac{\partial G_{2}^{(1)} (1,2)}{\partial v_{1}} \, F (3) \, U' (\theta_{1} \!-\! \theta_{3}) 
\nonumber
\\
& \, - \mu \frac{\partial G_{2}^{(1)} (1,3)}{\partial v_{1}} \, F (2) \, U' (\theta_{1} \!-\! \theta_{2})
\nonumber
\\
& \,  \bigg]_{(1,2,3)} = 0
\label{short_evol_G3}
\end{align}
All together, Eqs.~\eqref{short_evol_F},~\eqref{short_evol_G21},~\eqref{short_evol_G22},
and~\eqref{short_evol_G3} form the starting point to derive
the kinetic Eq.~\eqref{final_kinetic},
as we describe in Appendix~\ref{sec:Derivation}.

\section{Deriving the kinetic equation}
\label{sec:Derivation}

In this Appendix, we detail the protocol followed
to obtain the ${ 1/N^{2} }$ kinetic equation
put forward in Eq.~\eqref{final_kinetic},
following an approach similar to~\cite{RochaFilho2014}.
Here, we only present the overall approach
and the key steps, while the detailed effective (and cumbsersome)
computations were performed using symbolic calculations
in \texttt{Mathematica}, as detailed in~\cite{MMA}.
From the technical point of view, the main difficulty
is to deal, without mistake, with the large number of terms
that appear in the successive resolutions of the evolution equations,
hence the need for a numerical implementation of this calculation.

Luckily, the four truncated evolution equations,
Eqs.~\eqref{short_evol_F},~\eqref{short_evol_G21},~\eqref{short_evol_G22},
and~\eqref{short_evol_G3} form a closed and well-posed hierarchy
of coupled partial differential equations.
In particular, owing to the absence of any collective effects,
that would require for the explicit characterisation of the system's linear response,
the evolution equations can easily be solved in sequence.
The first step is to solve for the time evolution of ${ G_{2}^{(1)} (1,2) (t) }$,
as governed by Eq.~\eqref{short_evol_G21}.
This explicit solution may then be used as a (time-dependent) source term in Eq.~\eqref{short_evol_G3} to obtain the time evolution of ${ G_{3} (t) }$.
This function can then be used as a (time-dependent) source term in Eq.~\eqref{short_evol_G22}
to derive the time evolution of ${ G_{2}^{(2)} (t) }$.

In each of these three steps, we rely on two main assumptions:
(i) Bogoliubov's ansatz, so that we may take ${ F (1  , t) = \text{cst.} }$
when solving for the time-evolution of a correlation function;
(ii) we neglect the transients associated with any specific
initial conditions in the system's correlations,
i.e.\ we solve these differential equations with the initial conditions
${ G_{2}^{(1)} (t \!=\! 0) \!=\! G_{3} (t \!=\! 0) \!=\! G_{2}^{(2)} (t \!=\! 0) \!=\! 0 }$.
Moreover, in order to easily deal with phase mixing terms
of the form ${ v_{1} \partial G_{2}^{(1)} / \partial \theta_{1} }$, 
we perform Fourier developments of all the correlation functions
w.r.t.\ their $\theta$-dependence.
Similarly, the interaction potential is also expanded in its Fourier harmonics.
As imposed by Eq.~\eqref{def_U}, in the present case of the \HMF\ model,
the interaction potential takes the simple form ${ U (\theta) \!=\! \sum_{k = \pm 1} \!\tfrac{- U_{0}}{2} \re^{\ri k \theta} }$,
so that only the harmonics ${ k \!=\! \pm 1 }$ can support the interaction,
which offers a drastic reduction in the total number of resonant terms
that can contribute to the system's dynamics.

Following these three successive resolutions,
we now have at our disposal an explicit solution
for the time dependence of ${ G_{2}^{(2)} (t) }$.
Owing to Bogoliubov's ansatz,
we may then consider the limit ${ t \to + \infty }$ of that expression,
in order to obtain the asymptotic behaviour of $G_{2}^{(2)}$,
and inject it in Eq.~\eqref{short_evol_F}
to obtain the closed ${ 1/N^{2} }$ collision operator driving the long-term evolution
of ${ \partial F / \partial t }$.
At this stage, a typical time integral appearing in the expression
of ${ G_{2}^{(2)} (t) }$ takes the form
\begin{align}
\!\! \int_{0}^{t} \!\! \rd t_{1} \, \re^{\ri (t - t_{1}) \omega_{1}} & \, \frac{\partial }{\partial v_{1}} \bigg[ \!\! \int_{0}^{t_{1}} \!\! \rd t_{2} \, \re^{- \ri (t_{1} - t_{2}) \omega_{2}}
\nonumber
\\
\times & \, \frac{\partial }{\partial v_{2}} \bigg\{ \!\! \int_{0}^{t_{2}} \!\! \rd t_{3} \, \re^{- \ri (t_{2} - t_{3}) \omega_{3}} \bigg\} \bigg] ,
\label{shape_triple_integral}
\end{align}
where the frequencies $\omega_{1}$, $\omega_{2}$, and $\omega_{3}$
are some linear functions of the velocities $v_{1}$, $v_{2}$, and $v_{3}$,
i.e.\ the resonances involved in the dynamics,
while some additional gradients w.r.t.\ the velocities
can get intertwined with the time integrals.
Now, our goal is to estimate the asymptotic limit
${ t \to + \infty }$ of that expression in order to estimate the collision operator
driving the dynamics of ${ \partial F / \partial t }$.
To do so, we use the asymptotic formula
\begin{equation}
\lim\limits_{t \to + \infty} \!\! \int_{0}^{t} \!\! \rd t_{1} \, \re^{- \ri (t - t_{1}) \omega_{1}} = \pi \deltaD (\omega_{1}) - \ri \mP \bigg( \frac{1}{\omega_{1}} \bigg) ,
\label{formula_resonance}
\end{equation}
with ${ \deltaD (\omega) }$ the Dirac delta, and ${ \mP (1 / \omega) }$
the Cauchy principal value~(see the expression for ${ \delta_{+} (x) }$
in Eq.~{(6.40)} of~\cite{Balescu1997}).
For nested integrals as in Eq.~\eqref{shape_triple_integral},
we apply consecutively the formula from Eq.~\eqref{formula_resonance}\footnote{One
could be concerned by the nested bounds from the three successive integrals of Eq.~\eqref{shape_triple_integral}.
Even if one has
${ \!\!\int_{0}^{t} \!\! \rd t_{1} \!\! \int_{0}^{t_{1}} \!\! \rd t_{2} \!\! \int_{0}^{t_{2}} \!\! \rd t_{3} \!=\! \tfrac{1}{6} \!\! \int_{0}^{t} \!\! \rd t_{1} \!\! \int_{0}^{t} \!\! \rd t_{2} \!\! \int_{0}^{t} \!\! \rd t_{3} }$,
when applying successively the formula from Eq.~\eqref{formula_resonance},
the ${ \tfrac{1}{6} }$ volume prefactor does not have to be accounted for.}.
Doing so, one still prevents for now the evalutation of the gradients
w.r.t.\ the velocities,
so that such gradients would only act on the Dirac deltas
and the Cauchy principal values.

Once all the time integrals have been replaced by their asymptotic behaviours,
the derived kinetic equation takes the form
\begin{align}
\frac{\partial F (v_{1})}{\partial t} = & \, \frac{\pi^{3}}{2} U_{0}^{4} \mu^{2} \frac{\partial }{\partial v_{1}} \bigg[ \!\! \int \!\! \rd v_{2} \rd v_{3}
\nonumber
\\
\times \bigg\{ & \, \deltaD (2 v_{1} - v_{2} - v_{3}) \, \bigg[ \mP \, \mPp \, \KI + \mP \, \mPpp \, \KII  \bigg] 
\nonumber
\\
- & \, ( v_{1} \leftrightarrow v_{2} ) \bigg\} \bigg] ,
\label{ugly_kinetic}
\end{align}
where we introduced the shortening notations
${ \mP = \mP (\tfrac{1}{v_{1} - v_{2}}) }$,
${ \mPp = \mPp (\tfrac{1}{v_{1} - v_{2}}) }$,
and ${ \mPpp = \mPpp (\tfrac{1}{v_{1} - v_{2}}) }$.
When making the substitution ${ ( v_{1} \leftrightarrow v_{2} ) }$ in Eq.~\eqref{ugly_kinetic},
it is important to note that ${ (\mP , \mPp , \mPpp) \to (- \mP , \mPp , -\mPpp) }$.
Finally, in Eq.~\eqref{ugly_kinetic}, we also introduced the differential
operators
\begin{align}
\KI = \bigg[  & \, 2 \, \frac{\partial }{\partial v_{2}} \, \frac{\partial }{\partial v_{3}} - 3 \, \frac{\partial }{\partial v_{1}} \, \frac{\partial }{\partial v_{3}} 
\label{def_KI}
\\
& \, + 2 \, \frac{\partial^{2} }{\partial v_{3}^{2}} + \frac{\partial }{\partial v_{1}} \, \frac{\partial }{\partial v_{2}} - 2 \, \frac{\partial^{2} }{\partial v_{1}^{2}} \bigg] \, F (v_{1}) \, F (v_{2}) \, F (v_{3}) ,
\nonumber
\end{align}
and
\begin{equation}
\KII = \bigg[ \frac{\partial }{\partial v_{3}} + \frac{\partial }{\partial v_{2}} - 2 \, \frac{\partial }{\partial v_{1}} \bigg] \, F (v_{1}) \, F (v_{2}) \, F (v_{3}) .
\label{def_KII}
\end{equation}
We note that Eq.~\eqref{ugly_kinetic} is almost
identical to the ${ 1/N^{2} }$ kinetic equation
already put forward in Eq.~{(23)} of~\cite{RochaFilho2014}
for the same physical system.
The differences are some corrections in the overall prefactor,
and the overall sign of the ${ (v_{1} \leftrightarrow v_{2}) }$ term.

Luckily, the result from Eq.~\eqref{def_KII} can be
significantly simplified,
by using integration by parts,
as well as the parity symmetries of the Dirac deltas,
the principal values and their derivatives,
leading to the final result from Eq.~\eqref{final_kinetic}.
The detailed steps for these calculations
can be found in~\cite{MMA}.
We briefly present them below for completeness.

The key step is to perform an integration by parts in Eq.~\eqref{def_KII}
w.r.t.\ the integration variable $v_{2}$, using the formula
\begin{equation}
\mP \, \mPpp = - \frac{\partial }{\partial v_{2}} \bigg[ \mP \, \mPp \bigg] - \big( \mPp \big)^{2} .
\label{integration_by_parts_formula}
\end{equation}
At this stage, the derivatives of the Dirac deltas that appear
are subsequently integrated using an integration by parts
w.r.t.\ the integration variable $v_{3}$, so that
\begin{equation}
\deltaDp (2 v_{1} - v_{2} - v_{3}) = - \frac{\partial }{\partial v_{3}} \bigg[ \deltaD (2 v_{1} - v_{2} - v_{3}) \bigg] ,
\label{integration_by_parts_Dirac}
\end{equation}
and similarly for ${ \deltaDp (2 v_{2} - v_{1} - v_{3}) }$.
Proceeding that way allows us not to create any higher order
derivatives of the Cauchy principal values.
The kinetic equation then becomes simpler, as it reads
\begin{align}
\frac{\partial F (v_{1})}{\partial t} = \frac{\pi^{3}}{2} U_{0}^{4} \mu^{2} & \, \frac{\partial }{\partial v_{1}} \bigg[ \!\! \int \!\! \rd v_{2} \rd v_{3}
\nonumber
\\
\times \bigg\{ & \, \deltaD (\bk_{1} \cdot \bvel) \, \mP \, \mPp \, \MI
\nonumber
\\
+ & \, \deltaD (\bk_{1} \cdot \bvel) \, \big( \mPp \big)^{2} \, \MII (\bk_{1}) 
\nonumber
\\
+ & \, \deltaD (\bk_{2} \cdot \bvel) \, \big( \mPp \big)^{2} \, \MII (\bk_{2})
 \bigg\} \bigg] .
\label{less_ugly_kinetic}
\end{align}
In that equation, we introduced the differential operators
\begin{align}
\MI = & \, \bigg[ -2 \frac{\partial^{2}}{\partial v_{1}^{2}} + \frac{\partial^{2}}{\partial v_{2}^{2}} + \frac{\partial^{2} }{\partial v_{3}^{2}} + 2 \frac{\partial }{\partial v_{2}} \frac{\partial }{\partial v_{3}}
\nonumber
\\
& \, - \frac{\partial }{\partial v_{1}} \frac{\partial }{\partial v_{2}} - \frac{\partial }{\partial v_{1}} \frac{\partial }{\partial v_{3}} \bigg] \, F (v_{1}) \, F (v_{2}) \, F (v_{3}) ,
\label{def_MI}
\end{align}
and
\begin{equation}
\MII (\bk) = \bigg( \bk \cdot \frac{\partial }{\partial \bvel} \bigg) \, F (v_{1}) \, F (v_{2}) \, F (v_{3}) ,
\label{def_MII}
\end{equation}
with the resonance vectors $\bk_{1}$ and $\bk_{2}$
already defined in Eq.~\eqref{def_bk}.

At this stage, we finally note that the term in ${ \mP \mPp }$
in Eq.~\eqref{less_ugly_kinetic} will not contribute to the dynamics.
Indeed, from Eq.~\eqref{def_MI}, we note that $\MI$ is invariant under the change
${ (v_{2} \leftrightarrow v_{3}) }$.
This symmetry can be leveraged to get rid of this term, as follows.
Owing to the presence of the double integral ${ \!\int\! \rd v_{2} \rd v_{3} }$,
one can perform the symmetrisation ${ (v_{2} \leftrightarrow v_{3}) }$ for that term.
This leaves the Dirac delta, ${ \deltaD (\bk_{1} \cdot \bvel) }$, invariant.
From that same resonance condition, we note that the arguments of the Cauchy principal values
are transformed as ${ \tfrac{1}{v_{1} - v_{2}} \!\to\! \tfrac{1}{v_{1} - v_{3}} \!=\! - \tfrac{1}{v_{1} - v_{2}} }$.
Given the parities of $\mP$ and $\mPp$,
we can therefore conclude that the term in ${ \mP \mPp }$
in Eq.~\eqref{less_ugly_kinetic} is antisymmetric under the change
${ (v_{2} \leftrightarrow v_{3}) }$ so that the overall contribution
of this term vanishes.
The last step of the calculation is finally
to perform the replacement
${ \!\int\! \rd v_{2} \, \big( \mPp \big)^{2} \!\to\! \mP \!\int\! \rd v_{2} /(v_{1} \!-\! v_{2})^{4} }$.
All in all, one finally obtains the closed kinetic equation
spelled out in Eq.~\eqref{final_kinetic}.

\section{Linear response theory}
\label{sec:LinearTheory}

As highlighted in Appendix~\ref{sec:Truncation},
in order to obtain the kinetic Eq.~\eqref{final_kinetic},
we had to neglect the contributions from collective effects
in the BBGKY evolution equations.
As a consequence, in order to test that kinetic equation,
it is essential to place ourselves in regimes where collective effects
are indeed unimportant.
Luckily the strength of the self-gravitating amplification can be directly
estimated by solving the linear response theory of the system.
This is what we briefly reproduce in that Appendix.

A systematic approach to perform this calculation
is to proceed by analogy starting from
the generic result regarding the linear stability analysis
of (inhomogeneous) long-range interacting systems.
Following Eq.~{(5.94)} of~\cite{BinneyTremaine2008}
(similarly Eq.~{(G3)} of~\cite{FouvryBarOr2018} and references therein),
a system's stability is governed by the response matrix
\begin{equation}
\hbM_{pq} (\omega) \!=\! 2 \pi \sum_{k} \! \int \! \rd J \, \frac{k \, \partial F / \partial J}{\omega - k \, \Omega (J)} \psi^{(p) *}_{k} (J) \psi^{(q)}_{k} (J) ,
\label{Fourier_M}
\end{equation}
where the angle-action coordinates are ${ (\theta , J) = (\theta , v) }$,
the orbital frequencies are ${ \Omega (J) = v }$.
A system is then linearly unstable if there exists a complex frequency
${ \omega = \omega_{0} \!+\! \ri \eta }$, with ${ \eta > 0 }$,
such that ${ \hbM (\omega) }$ has an eigenvalue equal to $1$.
The natural basis elements follow from the pairwise interaction from Eq.~\eqref{def_U},
that can be written under the separable form
\begin{align}
U (\theta_{1} - \theta_{2}) & \, = - \sum_{p = \pm 1} \psi^{(p)} (\theta_{1}) \, \psi^{(p) *} (\theta_{2}) ,
\nonumber
\\
\psi^{(p)} (\theta) & \, = \sqrt{U_{0} / 2} \,\, \re^{\ri p \theta} ,
\label{decomposition_U}
\end{align}
and it is straightforward to check that ${ (\psi^{(+)} , \psi^{(-)}) }$ indeed
form a biorthogonal basis,
as defined, e.g.\@, in Eq~{(G1)} of~\cite{FouvryBarOr2018}.
Similarly their Fourier transform w.r.t.\ the angle can easily be computed.
It is independent of the action $v$, and reads ${ \psi^{(p)}_{k} = \delta_{k}^{p} \sqrt{U_{0} / 2} }$.
Owing to this Kronecker delta, the ${ 2 \times 2 }$ response matrix
from Eq.~\eqref{Fourier_M} is then
diagonal.
We may finally introduce the susceptibility matrix (or dielectric function),
${ \beps = \bI - \hbM }$, that is also diagonal with the coefficients
\begin{equation}
\veps_{\pm} (\omega) = 1 \mp \pi \, U_{0} \!\! \int \!\! \rd v \, \frac{\partial F / \partial v}{\omega \mp v} .
\label{diag_eps}
\end{equation}
Let us emphasise that the result from Eq.~\eqref{diag_eps}
is identical to the result presented in Eq.~{(9)} of~\cite{ChavanisDelfini2009}.

Relying on the same de-dimensionalisation as in Eq.~\eqref{final_kinetic_ddim},
one can rewrite the susceptibility coefficient from Eq.~\eqref{diag_eps}
under the form
\begin{equation}
\veps_{\pm} (\oomega) = 1 \mp \frac{1}{Q} \!\! \int \!\! \rd u \, \frac{\partial \oF / \partial u}{\oomega \mp u} , 
\label{diag_eps_ddim}
\end{equation} 
with ${ \oomega = \omega \td }$ a dimensionless frequency.
We also recall that the dimensionless \PDF\@, ${ \oF (u) }$,
and the stability parameter, $Q$, were given in Eqs.~\eqref{def_oF} and~\eqref{def_Q}.

Luckily, it is possible to further precise our characterisation of the system's stability
for the range of \DFs\ that will be considered in this paper.
Following the generic shape of the test \PDF\ from Eq.~\eqref{def_GaussianAlpha},
let us therefore assume that the system's \DF\@, ${ F (v) }$, is single-humped,
i.e.\ it possesses a single maximum.
We will also assume that the \DF\@'s is even, so that this maximum is reached in ${ v = 0 }$.
Owing to this parity, in Eq.~\eqref{diag_eps_ddim}, we note that
${ \veps_{+} (\oomega) = \veps_{-} (\oomega) }$,
so that we may limit ourselves to only studying $\veps_{+} (\oomega)$.

Because the \DF\ is single-humped in ${ v = 0 }$,
following Nyquist's criterion
(see, e.g.\@, Section~{2.6} in~\cite{ChavanisDelfini2009}),
such a \DF\ is linearly stable if, and only if, ${ \veps_{+} (0) > 0 }$.
As a consequence, following Eq.~\eqref{diag_eps_ddim},
such a \DF\ is linearly stable if, and only if, one has
\begin{equation}
Q > \Qc = - \!\! \int \!\! \rd u \, \frac{\partial \oF / \partial u}{u} .
\label{stability_Q}
\end{equation}
Conveniently, it is straightforward to compute
these stability thresholds for the generic test \DF\ from Eq.~\eqref{def_GaussianAlpha}.
This \DF\ is parametrised by the power index $\alpha$,
such that ${ \alpha = 2 }$ corresponds to the Gaussian distribution.
One finds
\begin{align}
\Qc (\alpha = 2) & \, = 1 ,
\nonumber
\\
\Qc (\alpha = 4) & \, = \frac{4 \, \Gamma [3/4]^{2}}{\Gamma [1/4]^{2}} \simeq 0.46 .
\label{calc_Qc}
\end{align}
In summary, a homogeneous Gaussian \PDF\ (i.e.\ a homogeneous Boltzmann distribution)
is therefore linearly stable if, and only if, it satisfies ${ Q > 1 }$.

\begin{figure}
\begin{center}
\includegraphics[width=0.48\textwidth]{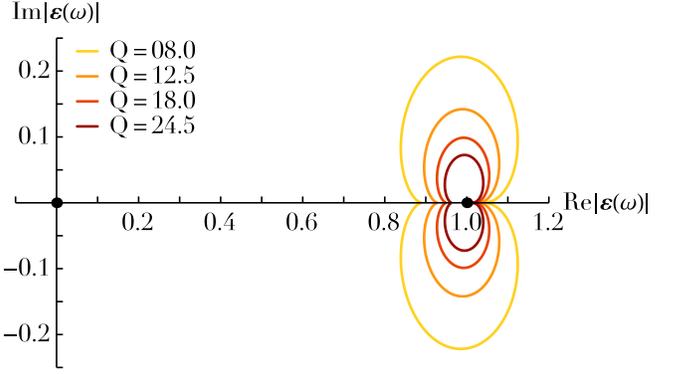}
\caption{Illustration of the Nyquist contours ${ \omega \mapsto \det[\beps (\omega + \ri \!\times\! 10^{-6})] }$
for the \PDF\ from Eq.~\eqref{def_GaussianAlpha} with ${ \alpha = 4 }$,
and different dynamical temperatures.
None of these contours enclose the origin,
indicating that all these systems are linearly stable.
The larger is $Q$, the closer is the contour to the point ${ (1,0) }$,
the weaker are collective effects,
and the more one is in the applicability regime of Eq.~\eqref{final_kinetic}.
\label{fig:Nyquist}}
\end{center}
\end{figure}

For more generic \PDFs\@, 
the susceptibility coefficient from Eq.~\eqref{diag_eps_ddim}
cannot always be computed analytically.
One has to resort to numerical evaluations,
e.g.\@, following the method presented in Appendix~{D} of~\cite{FouvryBarOrChavanis2019}.
We illustrate this method in Fig.~\ref{fig:Nyquist}
by representing the Nyquist contours associated with the test \PDF\
from Eq.~\eqref{def_GaussianAlpha}.
As expected, we recover in Fig.~\ref{fig:Nyquist}
that the larger is $Q$,
the weaker is the system's susceptibility,
i.e.\ dynamically hot systems are less efficient at amplifying
perturbations through collective effects.
Figure~\ref{fig:Nyquist} also illustrates
that the test case considered in Fig.~\ref{fig:Flux}
is sufficiently hot for the system to be linearly stable.
In addition, collective effects should prove sufficiently negligible
for the kinetic Eq.~\eqref{final_kinetic}
to be in its applicability regime.

\section{Numerical applications}
\label{sec:NBody}

In this Appendix, for completeness, we briefly present our numerical approach
to perform $N$-body simulations of the \HMF\ model.
Following Eq.~\eqref{total_H_HMF}, the specific Hamiltonian of a test particle embedded in that system reads
\begin{equation}
H_{\rt} (\theta_{\rt} , v_{\rt}) = \frac{v_{\rt}^{2}}{2} - M_{x} (t) \, \cos (\theta_{\rt}) - M_{y} (t) \, \sin (\theta_{\rt}) ,
\label{def_Ht}
\end{equation}
where we introduced the system's instantaneous magnetisations as
${ M_{x} (t) \!=\! U_{0} \mu \sum_{i = 1}^{N} \cos (\theta_{i} (t)) }$,
as well as
${ M_{y} (t) \!=\! U_{0} \mu \sum_{i = 1}^{N} \sin (\theta_{i} (t)) }$.
Two important remarks should be made w.r.t.\ Eq.~\eqref{def_Ht}.
First, because the pairwise interaction does not diverge at zero angular separation, the test Hamiltonian from Eq.~\eqref{def_Ht} and the associated
evolution equations can also be used to obtain correct evolution equations
for each of the system's particles, treating the magnetisations as external,
i.e.\ not taking any derivatives of it.
Second, as the magnetisations involve a sum over the $N$ particles,
they should be interpreted as global and shared quantities, that need to be computed only once for each timestep.
This allows for the computational complexity of integrating for one timestep
to scale like ${ \mO (N) }$, rather than ${ \mO (N^{2}) }$ as in the usual $N$-body problem.

To compute the velocity fields at a given time, we proceed as follows:
(i) we compute and store ${ (\cos , \sin) }$ for all particles;
(ii) we reduce these quantities to compute the instantaneous magnetisations ${ (M_{x} , M_{y}) }$;
(iii) we compute the velocity fields ${ \rd \theta_{i} / \rd t }$ and ${ \rd v_{i} / \rd t }$.
Owing to the fact that the Hamiltonian from Eq.~\eqref{def_Ht} is separable,
particles are then advanced using a fourth-order symplectic integrator
(see Eq.~{(4.6)} in~\citep{Yoshida1990}).
The numerical simulations presented in Section~\ref{sec:Applications}
were all performed using an integration timestep
equal to ${ \delta t = 1/(2 \sigma) }$ that guaranteed a relative error
in the total energy of the order of $10^{-7}$.

Following~\citep{NardonPianca2009}, the initial distribution of the system
is taken to be a generalised Gaussian distribution.
For a given index $\alpha$ and velocity dispersion $\sigma$, its \PDF\ reads
\begin{align}
P(u) & \, = \frac{\alpha}{2} \frac{A (\alpha , \sigma)}{\Gamma (1/\alpha)} \, \exp \!\big[ - (A (\alpha , \sigma) \, |u|)^{\alpha} \big] , 
\nonumber
\\
A (\alpha , \sigma) & \, = \frac{1}{\sigma} \, \bigg( \frac{\Gamma (3 / \alpha)}{\Gamma (1/\alpha)} \bigg)^{1/2} .
\label{def_GaussianAlpha}
\end{align}
This \PDF\ satisfies the normalisation condition
${ \!\int\! \rd u \, P(u) = 1 }$, is of zero mean and variance $\sigma^{2}$.
For ${ \alpha = 2 }$, this corresponds to the Gaussian distribution,
while larger values of $\alpha$ are associated with less peaked distributions.
Luckily, the \PDF\ from Eq.~\eqref{def_GaussianAlpha}
can also easily be sampled~(see Eq.~{(9)} in~\cite{NardonPianca2009}).

To measure fluxes, as in Figs.~\ref{fig:Flux} and~\ref{fig:Nscaling},
we proceeded as follows.
For each setup, we perform $\Nreal$ realisations,
only changing the initial conditions.
The dimensionless velocity space, ${ u \in [-3 , 3] }$,
is truncated in $50$ equal size bins.
For each realisation, each velocity bin location, and each timestep,
we compute the proportion of particles left to that location,
subsequently averaged over all the available realisations.
For each velocity bin, the associated time series are then fitted
with a linear time-dependence, whose slope is the local diffusion flux, ${ \omF (u , t \!=\! 0) }$.
To estimate the associated measurement errors, we follow the exact same approach
for $\Nboot$ measurements,
except that the sample of $\Nreal$ realisations over which the ensemble average
is performed allows for repetitions, i.e. for the same realisations to be used more than once.
The measurement is then given by the median value, while the errors are given
by the ${16\%}$ and ${84\%}$ confidence levels.
In Figs.~\ref{fig:Relaxation} and~\ref{fig:Flux}, we used the values ${ N = 10^{3} }$, ${ \Nreal \!=\! \Nboot \!=\! 10^{3} }$,
and simulated the systems up to ${ t = 5 \!\times\! 10^{8} \td }$.
With our implementation, running one such realisation
asked for about ${18\mathrm{h}}$ of computation on a single core.
In Fig.~\ref{fig:Nscaling}, we used the values ${ 0.5 \!\times\! 10^{3} \!\leq\! N \!\leq\! 1.7 \!\times\! 10^{3} }$,
${ \Nreal \!=\! \Nboot \!=\! 224 }$, and simulated the systems up to ${ t = 5 \!\times\! 10^{8} \td }$.

\end{document}